\documentclass[12pt]{iopart}
\usepackage{amssymb,graphicx}
\usepackage{color}
\usepackage{todonotes}
\usepackage{mcite}
\usepackage{cite}
\usepackage{braket}
\usepackage{iopams}
\usepackage{algorithm}
\usepackage{algpseudocode}

\definecolor{TealBlue}{rgb}{0.1,0.5,0.9}
\definecolor{Orange}{rgb}{0.8,0.3,0.0}

\definecolor{blue2}{rgb}{0.0, 0.3, 0.7}
\usepackage[colorlinks = true, citecolor = blue2, linkcolor = blue2, urlcolor=blue2]{hyperref}

\begin{document}

\title{A reinforcement learning approach to rare trajectory sampling}

\author{Dominic C. Rose\footnote{dominic.rose1@nottingham.ac.uk, corresponding author}, Jamie F. Mair\footnote{ppyjm13@nottingham.ac.uk} and Juan P. Garrahan\footnote{juan.garrahan@nottingham.ac.uk}}
\address{School of Physics and Astronomy 
\\ and \\
Centre for the Mathematics and Theoretical Physics of Quantum Non-Equilibrium Systems, 
\\
 University of Nottingham,  Nottingham NG7 2RD, United Kingdom}

\begin{abstract}
Very often when studying non-equilibrium systems one is interested in analysing dynamical behaviour that occurs with very low probability, so called {\em rare events}. In practice, since rare events are by definition atypical, they are often difficult to access in a statistically significant way. What are required are strategies to ``make rare events typical'' so that they can be generated on demand. Here we present such a general approach to adaptively construct a dynamics that efficiently samples atypical events. We do so by exploiting the methods of {\em reinforcement learning} (RL), which refers to the set of machine learning techniques aimed at finding the optimal behaviour to maximise a reward associated with the dynamics. We consider the general perspective of dynamical trajectory ensembles, whereby rare events are described in terms of ensemble reweighting. By minimising the distance between a reweighted ensemble and that of a suitably parametrised controlled dynamics we arrive at a set of methods similar to those of RL to numerically approximate the optimal dynamics that realises the rare behaviour of interest. As simple illustrations we consider in detail the problem of {\em excursions} of a random walker, for the case of rare events with a finite time horizon; and the problem of a studying current statistics of a particle hopping in a ring geometry,  
for the case of an infinite time horizon. We discuss natural extensions of the ideas presented here, including to continuous-time Markov systems, first passage time problems and non-Markovian dynamics.
\end{abstract}
\maketitle

\tableofcontents

\section{Introduction}

In physics, chemistry and many areas of science it is often the case that one wishes to study systems with dynamics which are highly variable and fluctuating, and where important information is contained in ``rare events'', meaning particular instances of the dynamics which are very far from typical. Since analytical study of the statistics of trajectories is almost always intractable beyond the simplest model systems one must resort to sampling trajectories numerically. The main challenge is how to access in an efficient manner the atypical trajectories that give rise to the rare events of interest \cite{Bolhuis2002,Garrahan2018}. 

A common problem is that of estimating the large deviation (LD) statistics \cite{Touchette2009} of time-extensive observables in systems with Markovian stochastic dynamics. This is difficult in general \cite{Giardina2006,Cerou2007,Lecomte2007b,Gorissen2009,Giardina2011,Nemoto2014,Nemoto2016,Nemoto2017,Nemoto2018,Ray2018,Ray2018b,Klymko2018,Ferre2018,Banuls2019,Helms2019,Jacobson2019,Ray2019,Helms2020} as such observables are concentrated around their average values which makes accessing the tails of their distributions an exponentially in time hard numerical task. In the dynamical LD context, several approaches have been developed which attempt to ameliorate the exponential scarcity of rare trajectories within the original dynamics, often based based either on population dynamics, such as cloning or splitting \cite{Giardina2006,Cerou2007,Lecomte2007b,Dean2009,Giardina2011,Carollo2019}, or on importance sampling in trajectory space, such as transition path sampling (TPS)
\cite{Bolhuis2002,Hedges2009}. 

Since rare events by definition are hard to obtain with the original dynamics of the system, a key approach is to find an alternative sampling dynamics that gives access to rare trajectories in an optimal manner \cite{Borkar2002,Borkar2003,Ahamed2006,Basu2008,Todorov2009,Chetrite2015,Jack2015a,Garrahan2016,Jack2019,Derrida2019,Derrida2019a,Dolezal2019,Oakes2020}. 
There is an intuitive similarity \cite{Gillman2020} in this search for an optimal sampling dynamics and the general problem of reinforcement learning (RL) \cite{Sutton2018}. Specifically, direct parametrisation of dynamics, such as the one done in the context above of trajectory sampling, is akin to policy gradient methods \cite{Williams1987,Williams1992} within RL. Exploring the connections between rare trajectory sampling and RL is the main aim of this paper. 

The use of RL methods in physics is of course a rapidly growing area. Examples include applications in quantum state preparation and quantum control \cite{Bukov2018,Bukov2018a,Fosel2018,Chen2019,Yao2020,Bolens2020}, quantum eigenstates \cite{AlbarranArriagada2020,Barr2020}, policy guided Monte Carlo simulations \cite{Bojesen2018}, and evolutionary RL for LDs \cite{Whitelam2019} and for thermodynamic control \cite{Beeler2019}. 

The key results and contributions of this paper are the following. 
(i) Using a generic formulation, which includes studying conditioned dynamics and cumulant generating functions as special cases, we demonstrate that the problem of optimizing a dynamics for sampling rare trajectories is identical to a form of regularized RL.
This connection both allows the adaptation of RL techniques to be used in sampling rare trajectories, and provides a new range of problems on which RL techniques can be tested and compared.
(ii) This form of regularized RL has not previously been considered using policy-gradient based techniques. We pedagogically present a range of such techniques for optimizing the sampling of rare trajectories. 
(iii) We review a small portion of the broad range of possible algorithms RL introduces through its connection with rare trajectory sampling. 
(iv) We specialize to the long-time limit, relevant to the large deviations of Markov chains, finding that the regularized RL algorithms automatically estimate the scaled cumulant generating function in the process of optimizing the dynamics.

The approach we present here has connections - but also important differences - to recent works exploring related ideas \cite{Todorov2007}, particularly in diffusive processes \cite{Kappen2012,Kappen2016,Das2019}. It is demonstrated throughout using problems based on random walkers.

The paper is organised as follows. In section \ref{background} we review the trajectory ensemble method in systems with stochastic dynamics, discuss their reweighting, and how rare trajectories relate directly to such reweightings.
In section \ref{dynamical_gradients} we pedagogically develop general methods for rare trajectory sampling based on RL, focusing on obtaining the optimal dynamics for finite problems. 
These methods are based on minimising expected likelihood, or a Kullback-Leibler (KL) divergence, and directly connect to maximum entropy RL and regularization \cite{Neu2017,Geist2019,Haarnoja2017,Haarnoja2018,Levine2018}. 
We illustrate our approach with the simple (and solvable) example of random walk excursions \cite{Majumdar2015}. 
We follow this in section \ref{sec:review} by reviewing a range of possible variations of these algorithms found in the RL literature, translated into our setting, which are made available by the connection between regularized RL and trajectory sampling.
Section \ref{LDs} extends the ideas of sections \ref{background} and \ref{dynamical_gradients} to the case of long times, viewed as an infinite horizon problem, establishing the connection to LD theory.
This connection implies these algorithms for optimizing the dynamics simultaneously provide an estimate for the scaled cumulant generating function, discussed in section \ref{sec:ld-scgf}. 
We conclude with section \ref{conclusion} outlining further extensions and possible adaptations of the methods presented here. 
This paper is intended to be the first in a series of works exploring connections between the physical and mathematical understanding of trajectory ensembles, and the computer science understanding of reinforcement learning.
Code produced to produce results for the examples shown in this paper is available on Github at \cite{Rose2020}.

\section{Formulation and applications}
\label{background}
We begin by introducing the formalism we use to describe trajectory ensembles, followed by a precise definition of the reweighted ensembles we consider.
We then discuss how two cases in which rare trajectories have a significant impact, conditioned ensembles and cumulant generating functions, can be viewed as studies of reweighted trajectories ensemble.
Finally, we discuss how our approach relates to -- and crucially, differs from -- the traditional formulation of reinforcement learning.

\subsection{Formalism and aim: trajectory ensembles and reweightings}
\label{conditioned_ensembles}
We consider a system evolving over time $t$ with state $x_t$.
For simplicity we consider a discrete time dynamics given by Markovian transition probabilities $P(x_{t}|x_{t-1})$, with $t$ taken to be a dimensionless integer denoting how many steps have occurred since the initial state.
However, in section \ref{generalizations} we discuss the simple extension to non-Markovian problems: for example, a simple modification is to make the dynamics time dependent, with transition probabilities $P(x_{t}|x_{t-1},t)$.

Trajectories consisting of sequences of states are labelled as
\begin{eqnarray}
\omega_{t_0}^T=\{x_t\}_{t_0}^T,
\end{eqnarray}
where $x_t$ is the state at time $t$, $t_0$ is the initial time and $T$ is the final time.
When $\omega$ appears multiple times in the same equation, we follow the convention that where their times overlap, they refer to the same states.
The probability of each trajectory is then given by
\begin{eqnarray}
	P\left(\omega_{0}^T\right)=\prod_{t=1}^TP(x_t|x_{t-1})P(x_0),
\end{eqnarray}
where $P(x_0)$ is the probability of a trajectory being initialized in the state $x_0$.
These trajectory probabilities define a trajectory ensemble that we will frequently refer to as the \textbf{original dynamics} $P$.
Throughout the paper we will make extensive use of expectation values over different trajectory ensembles, which we shall denote
\begin{eqnarray}
	\left\langle O\left(\omega_0^T\right)\right\rangle_P = \sum_{\omega_0^T}P\left(\omega_{0}^T\right)O\left(\omega_0^T\right),
\end{eqnarray}
where $O$ is some function of the trajectory and the subscript denotes the trajectory ensemble over which the expectation is taken.
We will also use conditional expectations over the future of a state
\begin{eqnarray}
	\left\langle O\left(\omega_{t}^T\right)\right\rangle_{P,X_t=x} = \frac{\sum_{\omega_0^T:x_t=x}P\left(\omega_{0}^T\right)O\left(\omega_{t}^T\right)}{\sum_{\omega_0^T:x_t=x}P\left(\omega_{0}^T\right)},
\end{eqnarray}
where $X_t$ denotes the random variable corresponding to the state at time $t$. 
Finally we will make use of the fact that the expectation of an expectation is simply the expected value: more specifically, we will use the identity
\begin{eqnarray}\label{eq:double-expectation}
	\left\langle f\left(\omega_t^T\right)g(x_t)\right\rangle_P=
	\left\langle \left\langle f\left(\omega_t^T\right)\right\rangle_{P,X_t=x_t}g(x_t)\right\rangle_P.
\end{eqnarray}

The problem we consider in this paper is finding a new dynamics which efficiently samples rare trajectories of some original Markovian dynamics $P$ as defined above.
In the next subsection we will provide examples showing many rare trajectory problems can be framed as the task of sampling a reweighting of the original trajectory ensemble.
As such, we will now define what we generally mean by a reweighted trajectory ensemble.
We will consider a weighting function which possesses a Markovian product structure: that is, the weight for each trajectory is given by
\begin{eqnarray}\label{eq:product-weighting}
	W\left(\omega_0^T\right)=\prod_{t=1}^TW(x_t,x_{t-1},t),
\end{eqnarray}
where
\begin{eqnarray}
	W(x_t,x_{t-1},t) \geq 0 \quad\forall\quad(x_t,x_{t-1},t).
\end{eqnarray}
This defines a reweighted trajectory ensemble as
\begin{eqnarray}\label{eq:reweighted-trajectory}
	P_W\left(\omega_0^T\right) = \frac{W\left(\omega_0^T\right)P\left(\omega_0^T\right)}{\left\langle W\left(\omega_0^T\right)\right\rangle_P}.
\end{eqnarray}
Our goal is then to find a new Markovian dynamics which generates a trajectory ensemble as close to this as possible, in a precise sense defined in terms of the Kullback-Leibler divergence in section \ref{dynamical_gradients}.
While it is not immediately clear from equation \eref{eq:reweighted-trajectory}, these trajectory probabilities can always be decomposed exactly into a set of time-dependent Markovian transition probabilities, as demonstrated in \hyperref[exact-excursions]{Appendix A}.
Conditions for when this is the case in diffusive systems have previously been studied under the moniker of penalizations in probability theory \cite{Roynette2009}.
However, for complex problems it will be difficult to calculate this \textbf{exact dynamics}.
It is for this reason that we present an approximate approach based on mapping the problem onto a regularized form of reinforcement learning.

We note here that, similar to how this approach extends naturally to a non-Markovian original dynamics, more general trajectory reweightings can be considered than the Markovian product structure of equation \eref{eq:product-weighting}.
For more general reweightings the exact dynamics which reproduces the reweighted ensemble is naturally non-Markovian, even if the original dynamics is not.
This is discussed further in section \ref{generalizations} and will be studied in future work.

\subsection{Applications: rare trajectories as reweighted ensembles}
We will now discuss how a variety of rare trajectory problems can be seen as a reweighting.
In this case the reweighted ensemble is difficult to study using simulations based on the original dynamics, necessitating the use of alternative sampling schemes such as cloning and TPS, and/or the construction of an adapted sampling dynamics
\cite{Giardina2006,Cerou2007,Lecomte2007b,Dean2009,Giardina2011,Carollo2019,Bolhuis2002,Hedges2009,Borkar2002,Borkar2003,Ahamed2006,Basu2008,Todorov2009,Chetrite2015,Jack2015a,Garrahan2016,Jack2019,Derrida2019,Derrida2019a,Dolezal2019,Oakes2020}.
Our work will supplement these by connecting the construction of an alternative sampling dynamics to reinforcement learning.

To make our discussion of applications concrete, we will use a simple model as a recurring example: a random walker.
That is, the original dynamics is that of a single particle hopping on a lattice, where the state $x$ takes integer values, with Markovian transition probabilities $P(x\pm1|x)=1/2$.
We will consider both infinite and periodic boundaries when we study rare events of this model in finite and long times, respectively.
The probability of each trajectory takes a particularly simple form, being just $P(\omega_{t}^T)=2^{-(T-t)}$.
We will consider a variety of rare event problems based on this model, related either to its instantaneous position $x$ or to an observable of the full trajectory, notably the area 
\begin{eqnarray}
A(\omega_{t}^T)=\sum_{t'=t}^Tx_{t'}.
\end{eqnarray}

\subsubsection{Conditioned dynamics}\label{sec:conditioned-dynamics}
The first class of problems we consider are those in which the trajectory ensemble is conditioned on some observation of the trajectory.
That is, a given some statement about the trajectory that is either true or false, we wish to consider only the subset of trajectories for which the statement is true.
Here the weight is simply a binary $W\left(\omega_0^T\right)=0$ if the statement is false, and $W\left(\omega_0^T\right)=1$ if the statement is true.
The resulting ensemble then consists of rare trajectories of the original dynamics if the probability of the condition being true is small.

For example, we may condition the trajectory ensemble of the random walker on ending in the state $x_T=0$, with an initial condition of $x_0=0$, often called a random walk bridge \cite{Majumdar2015}.
The weights for each transition would then be precisely defined as $W(x_T,x_{T-1},T)=\delta(x_T)$ and $W(x_t,x_{t-1},t)=1$ for $0<t<T$.
Such a trajectory is relatively rare in the original dynamics.
The probability of generating such a trajectory in the original dynamics is equal to the number of such trajectories, multiplied by their probability: the number of trajectories is simply the number of orderings of an equal number of up and down steps, resulting in
$P(x_T=0|x_0=0)\propto T^{-\frac{1}{2}}$.

A harder problem would be to retain the same constraint on the end, but additionally require $x_t\geq0$ for all $t$, known as random walk excursions \cite{Majumdar2015}. 
Using the step function $H(x_t)$, equal to zero for $x_t<0$ and one otherwise, the weights may then be written $W(x_T,x_{T-1},T)=\delta(x_T)$ and $W(x_t,x_{t-1},t)=H(x_t)$ for $0<t<T$.
As can be seen in \hyperref[exact-excursions]{Appendix A}, in this case the number of trajectories relates to Catalan numbers, with $P(x_T=0,x_t\geq0\forall t|x_0=0)\propto T^{-\frac{3}{2}}$.
Thus these excursions are substantially rarer than the bridges. 
Both excursions and bridges are have been studied extensively in a continuous time and space context of Brownian motion, see e.g. \cite{Majumdar2015}.

In our approach it will be necessary to have weights which are always non-zero.
As such, to consider conditioned problems we will first need to \textbf{soften} the weights, setting the trajectory weight to $1$ on correct trajectories and $<1$ on incorrect trajectories.
In particular, we can consider the weightings to be given by some measure $D$ which returns $0$ when the condition is true and is positive when the condition is false
\begin{eqnarray}\label{softened-constraint}
W\left(x_t,x_{t-1},t\right)=e^{-sD\left(x_t,x_{t-1},t\right)},
\end{eqnarray}
where $s$ is a parameter determining how heavily suppressed incorrect trajectories will be: in the limit $s\rightarrow\infty$, only correct trajectories remain, recovering the ensemble of the hard constraint.
For example, to recover a softened version of the random walk bridges or excursions, we may set
\begin{eqnarray}\label{soft-excursion-distance}
D\left(x_t,x_{t-1},t\right)=x_t^2\delta_{t,T}+b(1-H(x_t)),
\end{eqnarray}
where $b$ is a parameter, returning a softened bridge problem at $b=0$ and a softened excursion problem at $b>0$.

\subsubsection{Tilted ensembles and cummulant generating functions}
Suppose we wish to study the statistics of some time integrated observable
\begin{eqnarray}
	O\left(\omega_0^T\right)=\sum_{t=1}^To(x_t,x_{t-1},t).
\end{eqnarray}
This cound by done by considering conditioned ensembles for each of its possible values, however, this is often a difficult task even for a single value \cite{Derrida2019,Derrida2019a}.
While softened constraints are easier for individual values, annealing the constraint over a whole range of values could be computationally demanding.
A common solution is to instead consider the observables \textbf{cumulant generating function}, given by
\begin{eqnarray}
Z(s,T)=\left\langle e^{-sO\left(\omega_0^T\right)}\right\rangle_{P}.
\end{eqnarray}
This tells us about the observables statistics by generating the observables cumulants through its derivatives at zero
\begin{eqnarray}
\left. \frac{\partial^n Z}{\partial s^n}\right|_{s=0}=(-1)^n\left\langle O\left(\omega_0^T\right)^n\right\rangle_{P}.
\end{eqnarray}
For certain observables or values of $s$ substantially different from $0$, many trajectories may make negligible contribution to this expectation, i.e.\ it is dominated by rare events in the dynamics.
To sample these rare events more efficiently, we may thus seek a dynamics corresponding to an ensemble reweighted according to the value of this observable, that is
\begin{eqnarray}
	P_W\left(\omega_0^T\right) = \frac{e^{-sO\left(\omega_0^T\right)}P\left(\omega_0^T\right)}{Z(s,T)}.
\end{eqnarray}
often referred to as the biased or titled ensemble of trajectories.
For example, if we wanted to consider the statistics of the area, we would set $o(x_t,x_{t-1},t)=x_t$ and have
\begin{eqnarray}
	W\left(x_t,x_{t-1},t\right)=e^{-sx_t},
\end{eqnarray}
and thus
\begin{eqnarray}
W(\omega_0^T)=e^{-sA\left(\omega_0^T\right)}=\prod_{t=0}^Te^{-sx_t}.
\end{eqnarray}
	
A particular case of the above is the study of observables in the long time limit.
For appropriate observables in many models, the probability of a particular value takes a \textbf{large deviation} form \cite{Touchette2009}, finding
\begin{eqnarray}
P(O|T)\propto e^{-T\phi\left(\frac{O}{T}\right)},
\end{eqnarray}
where $\phi\left(\frac{O}{T}\right)$ is referred to as the rate function, describing the probability of the observable taking a particular value per unit time.
In these cases the cumulant generating function additionally has a simplified form, in terms of the scaled cumulant generating function (SCGF) $\theta(s)$
\begin{eqnarray}
Z(s,T)\propto e^{T\theta\left(s\right)}.
\end{eqnarray}
The SCGF $\theta(s)$ is thus often the aim of studies into the long-time statistics of time integrated observables, as it encodes the observables moments.
As we will see in section \ref{LDs}, such problems can be considered using a continuing form of RL.
In fact, a key result is that $\theta(s)$ ends up being directly related to the quantity we identify as our analogue of the return from RL, the precise quantity we will aim to maximize.
Our algorithms thus provide a two-for-one: they both find a dynamics which approximately generates the tilted ensemble, while simultaneously finding a variational approximation to $\theta(s)$.

\subsection{Relationship to standard reinforcement learning}
\label{MDP}

Here we will briefly describe how our problem relates to the standard approach to RL.
The aim of RL is to achieve some desired objective, by finding the best decisions or \textbf{actions} to make given some current information about the situation (the \textbf{state} of the environment) in which the objective must be achieved \cite{Sutton2018}.
Actions are chosen within each state according to a \textbf{policy}, which influences the transition to the next state.
The key ingredient of RL is inspired by behavioural psychology: the objective is encoded in a sequence of \textbf{rewards} received for each action in each state.
Formally these rewards are assigned real numbers, with the magnitude and sign defining how good or bad a decision is.
The resulting construction is referred to as a Markov Decision Process (MDP).
The goal of RL is then simply to maximize the sum of rewards -- the \textbf{return} -- received, thus making the best decisions to achieve the objective: this is done by optimizing the policy according to which actions are taken.

Our problem can be seen as a simplified form of RL in which each ``action'' precisely chooses the next state: we can therefore forgo the concept of actions and simply view the problem as choosing the best next state given the current state.
The dynamics is thus completely defined by the policy of how the next state is chosen.
To connect to RL, it thus remains to define the ``reward'' in our problem.
A natural suggestion for the return of each trajectory may be the log of the trajectories weight, which naturally produces a sum over terms associated to each transition
\begin{eqnarray}
	\ln W\left(\omega_0^T\right)=\sum_{t=1}^T\ln W(x_t,x_{t-1},t).
\end{eqnarray}
Maximising the return would thus result in a dynamics which produces trajectories of maximal weight.
However, while this is along the right lines, standard reinforcement learning tends to produce a deterministic policy: in this case, it would only produce trajectories with the maximum possible weight.
Our goal is to approximately reproduce the reweighted trajectory ensemble, producing each trajectory proportional to its weight.
This necessarily requires the transitions from each state to be probabilistic.
While there are ad-hoc approaches to making the learnt policy probabilistic, our key result is that there is in fact a natural way of framing our optimization problem as a \textbf{regularized} form of RL, based on the Kullback-Leibler divergence.
This regularized form is very similar to recent maximum-entropy RL techniques \cite{Haarnoja2017,Haarnoja2018,Levine2018} and other suggested approaches to regularizing RL \cite{Neu2017,Geist2019}, however, we are not aware of policy gradient techniques having been considered for the particular form of regularization our problem relates to.
This relation to regularized forms of RL will be discussed further in section \ref{sec:connect-reg-RL}.

The most significant tool this connection allows us to take from RL is that of value functions, which naturally emerge in a slightly modified form in this regularized setting.
These modified value functions satisfy a Bellman equation, as seen in section \ref{sec:actor-critic}.
Value focused approaches to RL often use this as a starting point, as do some policy focused approaches.
Equally, there exist many techniques, such as pure Monte-Carlo sampling, which make no use of Bellman equations in formulation or algorithmic solution \cite{Sutton2018}.
Further, they are not necessary in the initial introduction to policy-gradient techniques.

While important for our approach, we believe beginning our discussion by introducing both values and the Bellman equations they satisfy will serve to hide the simple connection between rare trajectory sampling and RL under further layers of abstraction.
Further, it would result in the rapid introduction of a range of concepts which are not common knowledge within the physics community.
As such, we choose to gradually introduce value functions and the Bellman equation as a natural tool in improving a gradient based approach, rather than a foundation, during the pedagogical development of the next section.

\section{Gradient optimization of rare finite-time trajectory sampling}
\label{dynamical_gradients}

In our approach, we seek to search through a space of \textbf{parameterized dynamics} $P_\theta(x_t|x_{t-1},t)$, conditional on the state and time, in order to make the trajectory ensemble it generates with probabilities given by
\begin{eqnarray}\label{current_trajectory_prob}
P_\theta(\omega_0^T)&=\prod_{t=1}^TP_\theta(x_t|x_{t-1},t)P(x_0),
\end{eqnarray}
as similar to the reweighted trajectory probabilities of equation \eref{eq:reweighted-trajectory} as possible.
Similarity is defined by the Kullback-Leibler (KL) divergence between the parameterized trajectory ensemble and the reweighted trajectory ensemble
\begin{eqnarray}\label{KL_divergence}
	D_{KL}(P_\theta|P_W)=\sum_{\omega_0^T}P_\theta\left(\omega_0^T\right)\ln\left(\frac{P_\theta\left(\omega_0^T\right)}{P_W\left(\omega_0^T\right)}\right)=\left\langle\ln\left(\frac{P_\theta\left(\omega_0^T\right)}{P_W\left(\omega_0^T\right)}\right)\right\rangle_{P_\theta},
\end{eqnarray}
taking value $0$ only when the trajectories distributions $P_\theta$ and $P_W$ are identical, a measure of similarity discussed in \cite{Chetrite2015} in the context of continuous time.
If these trajectory distributions agreed, we would refer to the parameterized dynamics $P_\theta(x_t|x_{t-1},t)$ as the \textbf{optimal dynamics}.
We take the expectation over the parameterized dynamics $P_\theta$, since this is precisely what we have access to, and can thus run simulations to sample it.
This differs from the approach recently considered for rare continuous-time diffusive trajectories in e.g. \cite{Kappen2016}, where the KL divergence is treated with the distributions reversed: the expectation is taken with respect to the reweighted distribution $P_W$, with expectation then calculated through importance sampling.
In principle, if $P_W$ is contained within the set of parameterized dynamics $P_\theta$, these KL divergences have the same minimum.
However, when this is not the case the two perspectives will differ in their optimal dynamics.

We will conduct our search through the space of dynamics by performing gradient descent optimization on the KL divergence \eref{KL_divergence}.
We thus require that the parameterized dynamics $P_\theta(x_t|x_{t-1},t)$ be differentiable with respect to the \textbf{weight} $\theta$.
We note that, to truly zero out the KL divergence, in general we would also have to parametrise and optimise the initial state distribution, as this will differ from its original form in the reweighted trajectory ensemble.
For simplicity, we will forgo including this initial distribution parametrisation and the resulting modifications to the algorithms, but their inclusion is a simple extension to what we will develop.

In the following sections, we will pedagogically demonstrate how to minimize this function efficiently through a line-search gradient descent based approach, following estimates of the gradient of equation \eref{KL_divergence}.
Similar to the policy gradient algorithms of RL, and thus referred to as dynamical gradient algorithms in the physical context, the resulting methods are very similar in structure to those found in maximum-entropy reinforcement learning \cite{Haarnoja2017,Haarnoja2018,Levine2018}, and closely related to current research in regularized MDPs \cite{Neu2017,Geist2019}.
Following an analogous development to that of \cite{Sutton2018}, we begin with a simple Monte Carlo sampling based algorithm closely related to \cite{Das2019}.
We then introduce an additional function approximation for the ``value'' of each state, used to guide the dynamical gradient first as a comparative baseline, and then as a bootstrapping estimate, leading to a so-called ``actor-critic'' algorithm.
In particular, our use of a value function to guide the optimization of the dynamics is a first in approaches focused on trajectory sampling: this provides a key example of the techniques that can be used due to our connection between trajectory sampling problems and RL.
We will not provide proofs of convergence or quality of converged results of the proposed algorithms in this work, however, we will apply several algorithms to a toy model, and reference theoretical results for similar RL algorithms throughout the section.

\subsection{Modifying transitions according to futures experienced: Monte Carlo returns}
First, for clarity, we rewrite the normalization factor, or ``partition function'', as
\begin{eqnarray}
Z=\left\langle W\left(\omega_{0}^T\right)\right\rangle_P.
\end{eqnarray}
Substituting the definitions of the parameterized trajectory probability \eref{current_trajectory_prob} and reweighted trajectory probability \eref{eq:reweighted-trajectory} into the KL divergence \eref{KL_divergence}, we have
\begin{eqnarray}\label{monte_carlo_dynamics_loss}
\fl D_{KL}(P_\theta|P_W)&=\left\langle\sum_{t=1}^T
 \ln\left(\frac{P_\theta(x_t|x_{t-1},t)}{P(x_t|x_{t-1})}\right)
-\sum_{t=1}^T\ln W(x_t,x_{t-1},t)
+\ln Z\right\rangle_{P_\theta}\nonumber\\
&=-\left\langle R(\omega_0^T)\right\rangle_{P_\theta}
+\ln Z
\end{eqnarray}
where we have defined the return $R$ of a trajectory as
\begin{eqnarray}
R(\omega_0^T)=\sum_{t=1}^T\ln W(x_t,x_{t-1},t) -
\sum_{t=1}^T
\ln\left(\frac{P_\theta(x_t|x_{t-1},t)}{P(x_t|x_{t-1})}\right),
\end{eqnarray}
encoding the contribution of each trajectory to the divergence, weighted by the probability.
Clearly, minimization of the KL divergence is analogous to maximization of the expected value of this return, similar to the usual situation considered in RL.
However, this differs from standard RL in the explicit dependence on the parameterized dynamics.
As a result, in contrast to standard RL where the return associated to each trajectory constant, here the return for a given trajectory changes with the parameterized dynamics.
This is the situation more commonly considered in maximum-entropy RL \cite{Haarnoja2017,Haarnoja2018,Levine2018}, where the attempt to maximize a return corresponding purely to the contribution of the weights is regularized by simultaneously trying to maximize the entropy of the trajectory ensemble.
For us, maximizing the RL reward is replaced by maximising the log of the weighting, while maximising entropy is replaced by minimizing the KL divergence between the original (non-reweighted) trajectory ensemble and the ensemble of the parameterized dynamics, an objective closely connected to current research in regularized MDPs \cite{Neu2017,Geist2019}.

For further clarity, we split the return into parts associated to each time step: specifically, we define an overall reward associated to each transition and time as
\begin{eqnarray}\label{reward}
r(x_t,x_{t-1},t)=\ln W(x_t,x_{t-1},t) -
\ln\left(\frac{P_\theta(x_t|x_{t-1},t)}{P(x_t|x_{t-1})}\right),
\end{eqnarray}
containing both the weighting and KL divergence contributions, such that the return on subsets of the trajectory is given by
\begin{eqnarray}
R(\omega_{t-1}^{t'})=\sum_{t''=t}^{t'}r(x_{t''},x_{t''-1},t'').
\end{eqnarray}

To minimize we will follow gradient descent on this objective, calculating its derivative with respect to the parameters $\theta$: noting 
\begin{eqnarray}
\fl&\nabla_\theta P_\theta(\omega_0^T)
=\nabla_\theta\prod_{t=1}^TP_\theta(x_{t}|x_{t-1})P(x_0)
=P_\theta(\omega_0^T)\sum_{t=1}^{T}\nabla_\theta\ln P_\theta(x_{t}|x_{t-1},t),\\
\fl&\nabla_\theta R(\omega_0^T)
=-\sum_{t=1}^{T}\nabla_\theta\ln P_\theta(x_{t}|x_{t-1},t),
\end{eqnarray}
we have
\begin{eqnarray}\label{monte_carlo_dynamics_gradient}
\nabla_\theta D_{KL}(P_\theta|P_W)
&=-\left\langle\left[R(\omega_0^T)-1\right]\sum_{t=1}^{T}\nabla_\theta\ln P_\theta(x_{t}|x_{t-1},t)\right\rangle_{P_\theta}\nonumber\\
&=-\left\langle\sum_{t=1}^{T}R(\omega_{t-1}^T)\nabla_\theta\ln P_\theta(x_{t}|x_{t-1},t)\right\rangle_{P_\theta}
\end{eqnarray}
where in the second line, we have removed the factor of 1 and the return prior to the differentiated time step of each summand, since
\begin{eqnarray}\label{total_derivative}
\sum_{x_t}P_\theta(x_{t}|x_{t-1},t)\nabla_\theta\ln P_\theta(x_{t}|x_{t-1},t)=\nabla_\theta\sum_{x_t} P_\theta(x_{t}|x_{t-1},t)=0,
\end{eqnarray}
due to the normalization of $P_\theta(x_{t}|x_{t-1},t)$.
Written in terms of the return, this takes the exact same form as the negative of the usual policy gradient of RL \cite{Sutton2018}, albeit with a regularized return.

As we will see below, Eq.~\eref{monte_carlo_dynamics_gradient} forms the basis of algorithms we will consider, as it can be manipulated into a wide variety of useful forms.
However, as stated this already provides an immediate algorithmic approach.

The exact value of the gradient specified by the above equation will be impossible to calculate even for simple problems.
Instead, since it takes the form of an expectation over trajectories, we can use Monte Carlo sampling of trajectories to construct an estimate, against which we will update the weights, before repeating the process.
Suppose we sample a set of $N$ trajectories $\{(\omega_i)_0^T\}_{i=1}^N$ using the current $P_\theta$ dynamics, each with partial returns after the state $x_{t}^i$ of
\begin{eqnarray}
	R_{t-1}^i=R\left((\omega_i)_{t-1}^T\right).
\end{eqnarray}
We can construct an empirical estimate of the gradient as
\begin{eqnarray}\label{eq:empirical-gradient}
\nabla_\theta D_{KL}(P_\theta|P_W)
&\approx-\frac{1}{N}\sum_{i=1}^N\left[\sum_{t=1}^{T}R_{t-1}^i\nabla_\theta\ln P_\theta(x^i_{t}|x^i_{t-1},t)\right].
\end{eqnarray}
We then update the weights by moving a short distance against the gradient, in order to reduce the KL divergence according to this estimate, as
\begin{eqnarray}
\theta_{n+1}=\theta_{n}+\alpha_n\frac{1}{N}\sum_{i=1}^N\left[\sum_{t=1}^{T}R_{t-1}^i\nabla_\theta\ln P_\theta(x^i_{t}|x^i_{t-1},t)\right],
\end{eqnarray}
where $\alpha_n$ is the learning rate for step $n$.
The estimate \eref{eq:empirical-gradient} can be calculated iteratively as each trajectory is created, updating the current average each new trajectory until a desired number has been run to reduce memory requirements.
Alternatively, we may even choose to sample a single trajectory between each update
\begin{eqnarray}\label{single_trajectory_mc_update}
\theta_{n+1}=\theta_{n}+\alpha_n\sum_{t=1}^{T}R_{t-1}\nabla_\theta\ln P_\theta(x_{t}|x_{t-1},t).
\end{eqnarray}

\begin{figure}
	\begin{center}
		\includegraphics[width=0.7\linewidth]{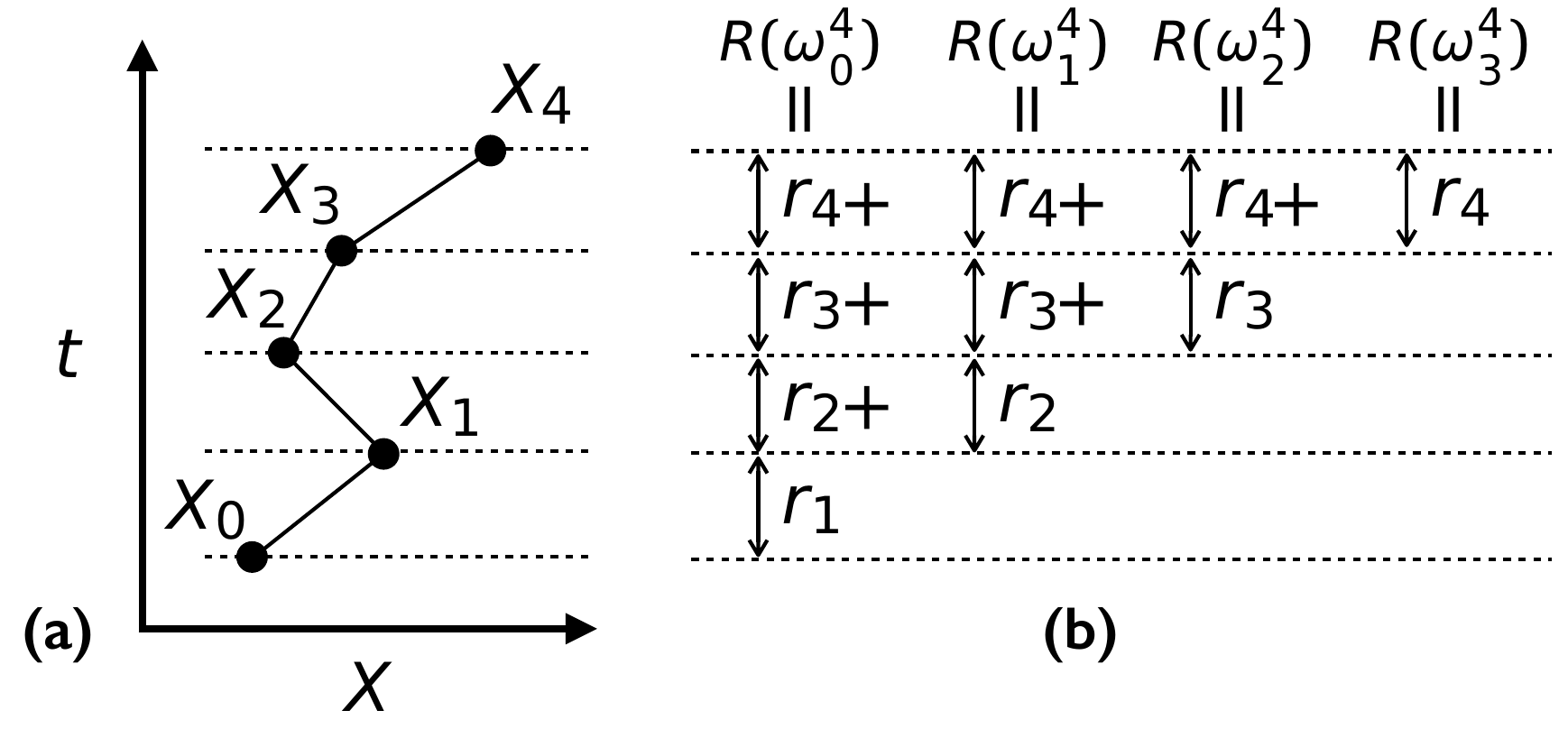}
	\end{center}
	\caption{\label{monte_carlo_sketch} Sketch of the information used in Monte Carlo return updates. (a) A simple sketch of an excursion, with space along the $x$ axis and time $t$ along the $y$ axis. (b) The information used to update the transitions originating from times $t=0,...,3$: the returns $R_t$ following each state $x_t$.}
\end{figure}

To gain an intuition for these updates, consider each term in the sum of equation \eref{single_trajectory_mc_update} individually, along the sample excursion trajectory of four steps sketched out in figure \ref{monte_carlo_sketch}(a).
The state $x_t$ at each time $t<T=4$ has an associated return $R_t$, given by the future rewards, see figure \ref{monte_carlo_sketch}(b).
Each term in the update \eref{single_trajectory_mc_update} then attempts to move the weights to increase or decrease the probability of the occurring transition, depending on the sign of the return: the size of the change is proportional to the magnitude of the resulting return.
A more rewarding future leads to a larger increase in transition probability, and vice-versa.
As many of these updates are committed, competing transitions (those for the same origin state) are then repeatedly enhanced or suppressed according to the resulting returns, leading to an eventual equilibration to a particular balance between the probabilities, depending on the returns that follow them.

Approaching this balance requires consideration of the learning rate $\alpha_n$: under ideal conditions on the function approximation and sampling, traditional RL convergence is expected provided the learning rate satisfies the requirements of the stochastic approximation
\begin{eqnarray}\label{learning_rates}
\sum_{n=0}^\infty\alpha_n=\infty, \qquad\qquad \sum_{n=0}^\infty|\alpha_n|^2=c,
\end{eqnarray}
where $c$ is any finite number \cite{Kushner2003,Borkar2008}.
However, convergence is only expected in the limit of infinite updates, and decaying learning rates can often slow learning.
In practice, learning rate which decrease (or even increase) for a short period at the start of learning, before becoming constant, may be beneficial \cite{Bertsekas1996, Sutton2018}.
For this algorithm, and standard RL algorithms without regularization, a constant learning rate will result in the weights fluctuating around a local minimum;
for the KL divergence regularized setting we consider, it in fact turns out that the components used in the algorithms introduced in later sections cause a decay of the gradient to zero, even for individual samples, as optimality is approached \cite{Nachum2017,Nachum2017a}.

More generically, both update rules described above fall under the umbrella of stochastic gradient descent, where noisy estimates of the gradient are used to update the parameters stochastically \cite{Borkar2008}.
The first of these updates is based on batches of trajectories, sometimes called mini-batches in the ML literatures, while the second is based on single samples.

The algorithm presented in this section is the simplest form of dynamical gradient algorithm, a regularized version of the classical REINFORCE algorithm \cite{Williams1987,Williams1992} based on return sampling, and as such we refer to this simply as KL regularized Monte Carlo returns.
For clarity, this algorithm is outlined below in algorithm \ref{monte_carlo_reinforce}.

\begin{algorithm}[h]
	\caption{KL regularized Monte Carlo returns}\label{monte_carlo_reinforce}
	\begin{algorithmic}[1]
		\State \textbf{inputs} dynamical approximation $P_\theta(x_t|x_{t-1},t)$
		\State \textbf{parameters} learning rate $\alpha_n$; total updates $N$
		\State \textbf{initialize} choose initial weights $\theta$, define iteration variables $n$ and $t$, total error $\delta_P$
		\State $n\gets0$
		\Repeat
		\State Generate a trajectory $\omega_0^T$ according to the dynamics given by $P_\theta(x_t|x_{t-1},t)$, with returns $R_t$ after each state $x_t$.
		\State $t\gets0$
		\State $\delta_P\gets0$
		\Repeat
		\State $\delta_P\gets\delta_P+R_{t-1}\nabla_\theta\ln P_\theta(x_{t}|x_{t-1},t)$
		\State $t\gets t+1$
		\Until{$t=T+1$}
		\State $\theta\gets\theta+\alpha_n\delta_P$
		\State $n\gets n+1$
		\Until{$n=N$}
	\end{algorithmic}
\end{algorithm}

\subsection{Comparing returns with past experiences: baselines and value functions}
A downside of this simple approach is the large potential variance in the return following a transition in each trajectory, which may provide an extremely noisy gradient from which to learn, resulting in slow convergence.
Fortunately, equation \eref{monte_carlo_dynamics_gradient} possesses an invariance which can be used to tame this variability.
Recalling how we used \eref{total_derivative} to remove the factor of one and the history of the return from \eref{monte_carlo_dynamics_gradient}, we may use this property to instead introduce any desired function of the past trajectory.
We introduce the \textbf{baseline} $b(x_t,t)$ as simply a function of the state and time, transforming \eref{monte_carlo_dynamics_gradient} into
\begin{eqnarray}
\fl \nabla_\theta D_{KL}(P_\theta|P_W)
&=-\left\langle\sum_{t=1}^{T}\left(R(\omega_t^T,x_{t-1})-b(x_{t-1},t-1)\right)\nabla_\theta\ln P_\theta(x_{t}|x_{t-1},t)\right\rangle_{P_\theta},
\end{eqnarray}
where the return following each transition is then contrasted with a baseline.

The choice of baseline can have a drastic impact on the variance of the gradient estimate, especially if we consider a small number of trajectories between updates.
A reasonable choice of baseline to minimize variance would simply be the average value of the return following a given state at a given time, the conditional expectation
\begin{eqnarray}\label{value_function}
V_{P_\theta}(x,t)=\left\langle R(\omega_{t}^T)\right\rangle_{P_\theta,X_t=x},
\end{eqnarray}
as this would minimize the variance of the baseline error
\begin{eqnarray}
\delta_b(\omega_{t-1}^T,t-1)=R(\omega_{t-1}^T)-b(x_{t-1},t-1),
\end{eqnarray}
and therefore might be expected to minimize the variance of the overall gradient estimate.
These \textbf{state values} encode the combined average weighting for the ensemble of sub-trajectories beginning from $x$ at time $t$, and KL divergence to the original dynamics of this sub-trajectory ensemble: the higher this value, the higher the average weighting and/or lower the KL divergence of this ensemble relative to that of the original dynamics.

The resulting gradient is given by
\begin{eqnarray}\label{monte_carlo_exact_value_baseline_loss}
\fl \nabla_\theta D_{KL}(P_\theta|P_W)
&=-\left\langle\sum_{t=1}^{T}\left(R(\omega_{t-1}^T)-V_{P_\theta}(x_{t-1},t-1)\right)\nabla_\theta\ln P_\theta(x_{t}|x_{t-1},t)\right\rangle_{P_\theta}.
\end{eqnarray}
Unfortunately, this is an ideal which can not be achieved: calculating the value for each state visited exactly is impossible in most problems of interest.
Instead, we introduce a second function approximation for the value function, $V_\psi(x_t,t)$, with weights $\psi\in\mathbb{R}^{d_V}$.
The exact error in each of the values provided by this function approximation is then given by
\begin{eqnarray}\label{state_value_loss}
L(\psi|x_t,t)=\frac{1}{2}\left(V_\psi(x_t,t)-V_{P_\theta}(x_t,t)\right)^2.
\end{eqnarray}

Even supposing we had an accurate result for the true value, we could not optimize these state-dependent loss functions one by one, as the resulting approximation would simply be overfitted on the last state optimized: instead, we must consider the states in unison.
However, we need not consider them with uniform weighting, and indeed each state will not be equally relevant to a given sampling dynamics and the rare event problem it is being optimized for.
The obvious choice for our aim is given by our current sampling dynamics: not only are we likely already using this to approximate the dynamical gradient, it will also prioritize the states which are most likely to occur in the current dynamics, and thus the most important to get accurate values for.
We thus sample states according to this dynamics, defining the loss function averaged over trajectories as
\begin{eqnarray}\label{return_value_loss}
L_V(\psi)&=\left\langle\frac{1}{2}\sum_{t=0}^{T-1}\left(V_\psi(x_t,t)-V_{P_\theta}(x_t,t)\right)^2\right\rangle_{P_\theta}
\end{eqnarray}
where the last time is neglected as the value is zero by definition.

Calculating the gradient of this loss, we have
\begin{eqnarray}
\nabla_\psi L_V(\psi)
&=\left\langle\sum_{t=0}^{T-1}\left(V_\psi(x_t,t)-V_{P_\theta}(x_t,t)\right)\nabla_\psi V_\psi(x_t,t)\right\rangle_{P_\theta},
\end{eqnarray}
giving a gradient in terms of the exact value similar to equation \eref{monte_carlo_exact_value_baseline_loss}: to get a target that can be evaluated we simply substitute the definition of the value \eref{value_function} and use \eref{eq:double-expectation} to find
\begin{eqnarray}\label{exact_value_gradient}
\nabla_\psi L_V(\psi)
&=-\left\langle\sum_{t=0}^{T-1}\left(R({\omega}_{t}^T)-V_\psi(x_t,t)\right)\nabla_\psi V_\psi(x_t,t)\right\rangle_{P_\theta}.
\end{eqnarray}

As with the dynamical gradient, to estimate the value loss functions gradient \eref{exact_value_gradient} we can simply sample one trajectory with states $x_t$ followed by returns $R_t$ leading to
\begin{eqnarray}\label{empirical_monte_carlo_value}
\nabla_\psi L_V(\psi)
\approx-\sum_{t=1}^{T}\left(R_{t-1}-V_\psi(x_{t-1},t-1)\right)\nabla_\psi V_\psi(x_{t-1},t-1).
\end{eqnarray}
Choosing a baseline $b(x_t,t)=V_\psi(x_t,t)$ then leads to an estimate for the policy gradient Eq. \eref{monte_carlo_exact_value_baseline_loss} for a given approximation of the value function, given by
\begin{eqnarray}\label{empirical_monte_carlo_value_policy_loss}
	\nabla_\theta D_{KL}(P_\theta|P_W)
	&\approx-\sum_{t=1}^{T}\left(R_{t-1}-V_{\psi}(x_{t-1},t-1)\right)\nabla_\theta\ln P_\theta(x_{t}|x_{t-1},t).
\end{eqnarray}
As in the previous section, we can readily construct empirical averages over multiple trajectories instead of considering single trajectories.

We can get some intuition for how the two approximations affect each other by considering how they affect each others loss functions and updates.
By construction, the dynamical gradient is on average independent of the baseline, and thus the optimal weights $\theta$ independent of the current values.
However, the better the value approximates the true values for the current policy, the smaller the variance in the updates and the faster the dynamics will converge.
We would thus desire the values to remain as accurate as possible to the current dynamics.
In contrast, the value loss function depends strongly on the dynamics: through the probability of each future trajectory, the priority given to each state, and the reward function itself.
The optimal value weights $\psi$ will thus depend strongly on the dynamics, however, for small changes in the policy we would expect a small change in the optimal value weights.
If the value function is reasonably accurate, a small change in the dynamics should thus only require a small number of updates to $\psi$ for it to again become accurate.

Accounting for these observations, there is some choice in the usage of updates given by equations \eref{empirical_monte_carlo_value} and \eref{empirical_monte_carlo_value_policy_loss}.
We could simply alternate updating the value function and the dynamics, leaving one fixed while the other changes.
This could range from letting the value function converge satisfactorily between updates to the dynamics, to simply alternating updates to the values and dynamics every trajectory.
Alternatively, we could use the same trajectory samples to simultaneously update both the dynamics and the values.
For a broader discussion of interleaving updates to the dynamics and values we refer to \cite{Sutton2018}, where it is discussed in particular under the terms asynchronous and generalized policy iteration.

\begin{figure}
	\begin{center}
		\includegraphics[width=0.8\linewidth]{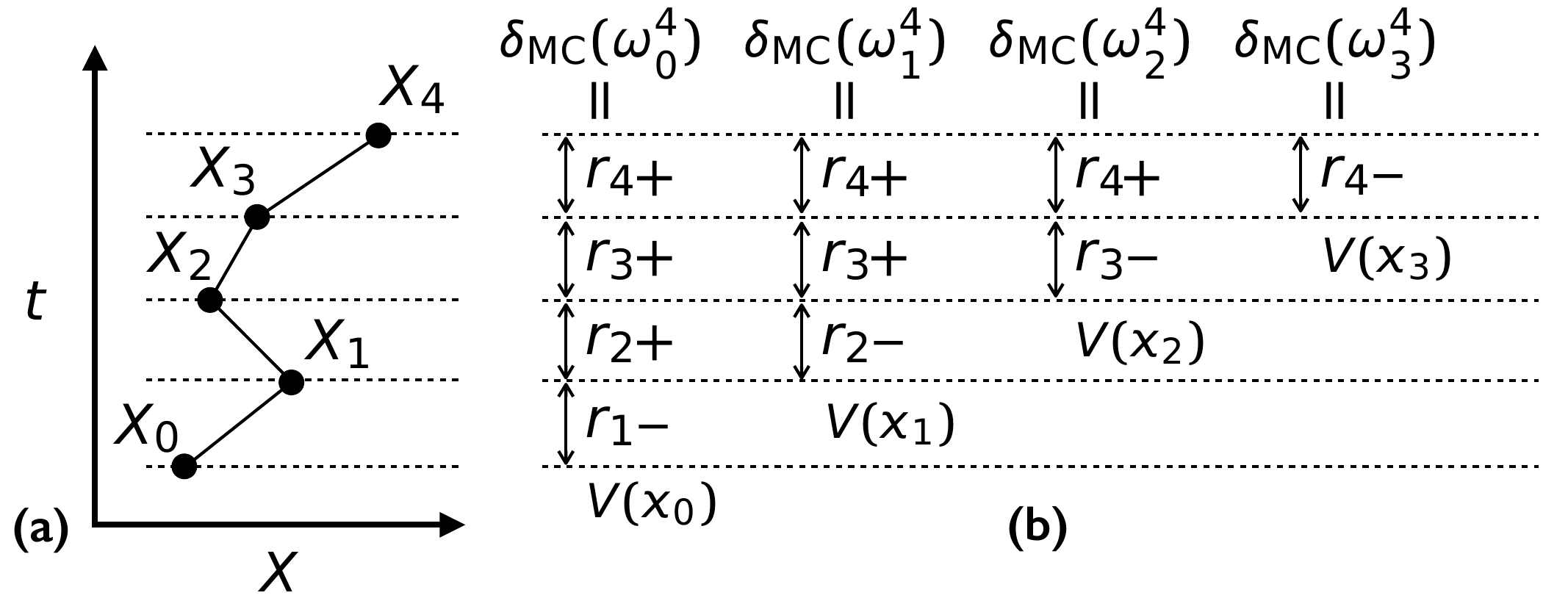}
	\end{center}
	\caption{\label{monte_carlo_value_sketch} Sketch of the information used in updates based on Monte Carlo returns with value baseline. (a) A simple sketch of an excursion, with space along the $x$ axis and time $t$ along the $y$ axis. (b) The information used to update the transitions originating from times $t=0,...,3$: the returns $R_t$ following each state $x_t$, contrasted with the value of that state $V_\psi(x_t)$.}
\end{figure}

The chosen scheme for updating both the values and dynamics in this double-learning scenario can have a significant affect on aspects of algorithm performance such as data efficiency, stability, convergence speed and bias in the final result.
For simplicity, we demonstrate using baselines with synchronous updates using a single trajectory for each update.
We refer to this as KL regularized Monte Carlo reinforce with a value baseline, due to its similarity to the Monte Carlo REINFORCE algorithm with a value function of reinforcement learning \cite{Sutton2018}.
Intuitively, for each trajectory we contrast the value of each state with the return following it, cf.\ figure \ref{monte_carlo_value_sketch}(b), aiming to increase both the probability of a transition and the value of a state if the return following it is greater than the value, and decrease them if the return is less.
We then conduct updates of the two weights $\theta$ and $\psi$ after every trajectory with learning rates $\alpha^\theta_n$ and $\alpha^\psi_n$ satisfying equations \eref{learning_rates}, in the directions suggested by the average of these return-value comparisons.
In practice, the efficiency of this algorithm is enhanced by noting that the factor multiplying the gradients in both updates takes the same form
\begin{eqnarray}\label{monte_carlo_error}
\delta_\mathrm{MC}(\omega_{t}^T,t)=R(\omega_{t}^T)-V_\psi(x_t,t),
\end{eqnarray}
which we refer to as the Monte Carlo value error.
It is outlined below in algorithm \ref{monte_carlo_reinforce_value_baseline}.

\begin{algorithm}[h]
	\caption{KL regularized Monte Carlo reinforce with value baseline}\label{monte_carlo_reinforce_value_baseline}
	\begin{algorithmic}[1]
		\State \textbf{inputs} dynamical approximation $P_\theta(x_t|x_{t-1},t)$, value approximation $V_\psi(x_t,t)$
		\State \textbf{parameters} learning rates $\alpha^\theta_n$, $\alpha^\psi_n$; total updates $N$
		\State \textbf{initialize} choose initial weights $\theta$ and $\psi$, define iteration variables $n$ and $t$, total errors $\delta_P$, $\delta_V$, individual error $\delta$
		\State $n\gets0$
		\Repeat
		\State Generate a trajectory $\omega_0^T$ according to the dynamics given by $P_\theta(x_t|x_{t-1},t)$, with returns $R_t$ after each state $x_t$.
		\State $t\gets0$
		\State $\delta_P\gets0$
		\State $\delta_V\gets0$
		\Repeat
		\State $\delta\gets R_{t}-V_\psi(x_{t},t)$
		\State $\delta_P\gets\delta_P+\delta\nabla_\theta\ln P_\theta(x_{t+1}|x_{t},t+1)$
		\State $\delta_V\gets\delta_V+\delta\nabla_\psi V_\psi(x_{t},t)$
		\State $t\gets t+1$
		\Until{$t=T$}
		\State $\theta\gets\theta+\alpha^\theta_n\delta_P$
		\State $\psi\gets\psi+\alpha^\psi_n\delta_V$
		\State $n\gets n+1$
		\Until{$n=N$}
	\end{algorithmic}
\end{algorithm}

Value baselines in the standard REINFORCE algorithm were considered in the original works on the algorithm \cite{Williams1987,Williams1992}, but more recent work has proposed that alternative baselines may provide a lower variance in the Monte Carlo setting \cite{Greensmith2004,Dick2015}, suggesting possible modifications to the above approach to further improve convergence rates.
Despite this, for the algorithms we consider next, it appears that the value baseline may indeed be the best choice \cite{Bhatnagar2009}.

\subsection{Replacing returns with past experiences: temporal differences and actor-critic methods}\label{sec:actor-critic}
\begin{figure}
	\begin{center}
		\includegraphics[width=0.8\linewidth]{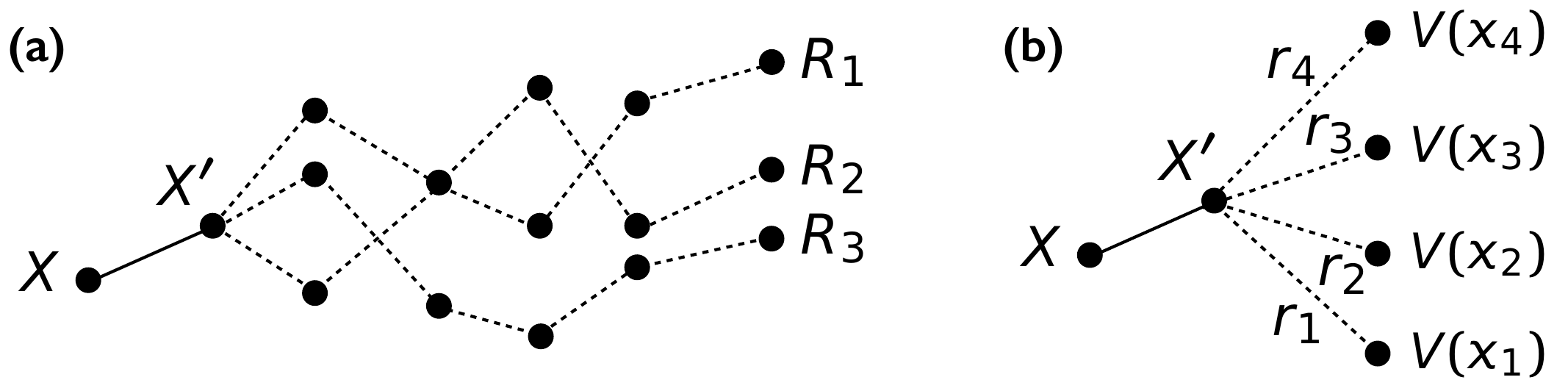}
	\end{center}
	\caption{\label{mc_ac_comparison_sketch} Comparison of updates used in Monte Carlo return updates and actor-critic updates. Whereas Monte Carlo returns (a) updates a transition $x\rightarrow x'$ according to the various possible returns following that transition, the 1-step actor-critic (b) update uses knowledge of only the reward during transition and estimates of the values of the states on either side of the transition.}
\end{figure}

The Monte Carlo error \eref{monte_carlo_error}, while better than the return alone, still possesses a relatively large variance if the remainder of the trajectory is long, the dynamics highly entropic and the weightings highly variable.
Further reduction of this variance would require an alternative to the return for contrast with the states values.
To this end, suppose we used many trajectory samples to construct an estimate of the gradient: transitions occurring multiple times will appear with their gradients multiplied by the average return following that transition, cf.\ figure \ref{mc_ac_comparison_sketch}(a).
Since the first reward is fixed by the transition, this average return would simply be the reward for that transition and the value of the state after transition.
This suggests that rather than contrasting the value of the state prior to the transition with the return of a whole trajectory, we could simply contrast the prior state value with the reward associated to that transition, and the estimated value of the resulting state built from past sampled trajectories.
If the estimated values are accurate, we would reasonably expect that on average this will result in the same gradients as using returns, cf.\ figure \ref{mc_ac_comparison_sketch}(b).

Unsurprisingly, this emerges naturally from the construction considered.
Beginning from equation \eref{monte_carlo_exact_value_baseline_loss} we immediately find
\begin{eqnarray}
	\fl \nabla_\theta D_{KL}(P_\theta|P_W)
	&=-\left\langle\sum_{t=1}^T\left(R(\omega_t^T)+r(x_t,x_{t-1},t)-V_{P_\theta}(x_{t-1},t-1)\right)\nabla_\theta\ln P_\theta(x_{t}|x_{t-1},t)\right\rangle_{P_\theta}\nonumber\\
	&=-\left\langle\sum_{t=1}^T\left(V_{P_\theta}(x_t,t)+r(x_t,x_{t-1},t)-V_{P_\theta}(x_{t-1},t-1)\right)\nabla_\theta\ln P_\theta(x_{t}|x_{t-1},t)\right\rangle_{P_\theta}\nonumber\\
\end{eqnarray}
where we have used \eref{eq:double-expectation} in the second line to replace the future return with the exact value.
Since we do not have access to the exact values of each state, we must approximate this expression using a value approximation.
Thus, defining a \textbf{temporal difference} (TD) error
\begin{eqnarray}\label{temporal_difference_error}
\delta_\mathrm{TD}(x_t,x_{t-1},t)=V_\psi(x_t,t)+r(x_t,x_{t-1},t)-V_\psi(x_{t-1},t-1),
\end{eqnarray}
so-called since it provides the difference between the value of the current state and the reward plus the value of the state at the next time, we have simply
\begin{eqnarray}\label{td_dynamical_gradient}
\nabla_\theta D_{KL}(P_\theta|P_W)
&\approx-\left\langle\sum_{t=1}^T\delta_\mathrm{TD}(x_t,x_{t-1},t)\nabla_\theta\ln P_\theta(x_{t}|x_{t-1},t)\right\rangle_{P_\theta},
\end{eqnarray}
which will be accurate whenever the value function is a good estimate for states which are commonly visited by the current dynamics $P_\theta$.
In reinforcement learning, such an approach is referred to as \textbf{actor-critic}, where the dynamics $P_\theta$ governing transitions would be the actor, while the value function $V_\psi$ judges the value of each state, playing the role of critic by informing the actor of whether a transition was good or bad.

For the critic, we could continue to use the Monte Carlo updates of the previous section, using the value function to construct approximate TD errors to update the dynamics.
However, the TD errors can also be used to update the critic itself, a process of updating estimates using estimates referred to as \textbf{bootstrapping}.
Beginning from equation \eref{exact_value_gradient}, following similar manipulation as that used to reach equation \eref{td_dynamical_gradient}, and substituting our approximation for the future value, we quickly arrive at
\begin{eqnarray}\label{td_value_gradient}
\nabla_\psi L_V(\psi)
\approx-\left\langle\sum_{t=0}^{T-1}\delta_{\mathrm{TD}}(x_{t+1},x_t,t)\nabla_\psi V_\psi(x_t,t)\right\rangle_{P_\theta},
\end{eqnarray}
analogous to the basic 1-step temporal difference value updates of RL \cite{Sutton1988}.
Clearly, for this to be an accurate approximation the value would already have to be accurate, thus suggesting this estimate would be poor when it matters: for weights $\psi$ which produce inaccurate values.
This brings into question how this gradient estimate could ever converge for an initially inaccurate set of weights.
Despite this, it often produces very successful results when used for updating the value weights in RL problems.

To understand why, we need to adopt a different perspective.
First we note that the exact value function satisfies a natural inductive definition
\begin{eqnarray}\label{bellman_equation}
V_{P_\theta}(x_t,t)=\left\langle V_{P_\theta}(x_{t+1},t+1)+r(x_{t+1},x_t,t)\right\rangle_{P_\theta,X_t=x_t},
\end{eqnarray}
commonly referred to as a Bellman equation, encoding the relationship between the value of state and other states visited in their immediate future.
As an alternative to our original choice of loss function \eref{return_value_loss}, using the returns along a trajectory, we could instead directly try to minimize the error in this equation for the approximation to the values.
That is, we could minimize the mean-squared Bellman error along a trajectory
\begin{eqnarray}\label{bellman_value_loss}
\fl L^{BM}_V(\psi)=\left\langle\frac{1}{2}\sum_{t=0}^{T-1}\left(\left\langle V_{P_\theta}(x_{t+1},t+1)+r(x_{t+1},x_t,t)\right\rangle_{P_\theta,X_t=x_t}-V_\psi(x_t,t)\right)^2\right\rangle_{P_\theta}.
\end{eqnarray}
Taking the derivative of this as is -- differentiating both the target expectation and the state sampled -- results in a complex gradient to calculate in general: this approach is addressed by so-called gradient-TD algorithms in the RL literature \cite{Sutton2009,Maei2009,Maei2011}, which have recently been extended to actor-critic methods \cite{Maei2018}.
While the unknown stochastic environment presents an additional issue requiring a double sampling of the transitions in that context, in our case the resulting gradient could alternatively be calculated exactly for each state visited, albeit at a substantial computational cost.

To jump from this alternative loss to the gradient of equation \eref{td_value_gradient} requires taking a slightly different view of the Bellman loss.
Suppose we instead minimize the distance between the value of each state and a target value predicted by the expectation on the right of equation \eref{bellman_equation} for the current weights.
That is, we keep the weights in the target expectation fixed and only differentiate the value of the state sampled from a trajectory.
Differentiating equation \eref{bellman_value_loss} with this fixed target and manipulating the expectations then leads directly to equation \eref{td_value_gradient}, but with a different interpretation: rather than approximating the gradient of the return based loss function, we are directly targeting an alternative prediction of the value based on the current estimated value of other states.
Such an approach is sometimes referred to as a ``semi-gradient'' method in the RL literature \cite{Sutton2018}, and has been seen to produce good results provided that the sampling of states is close to that of the dynamics the values are being estimated for, as discussed in more detail later.

\begin{figure}
	\begin{center}
		\includegraphics[width=0.8\linewidth]{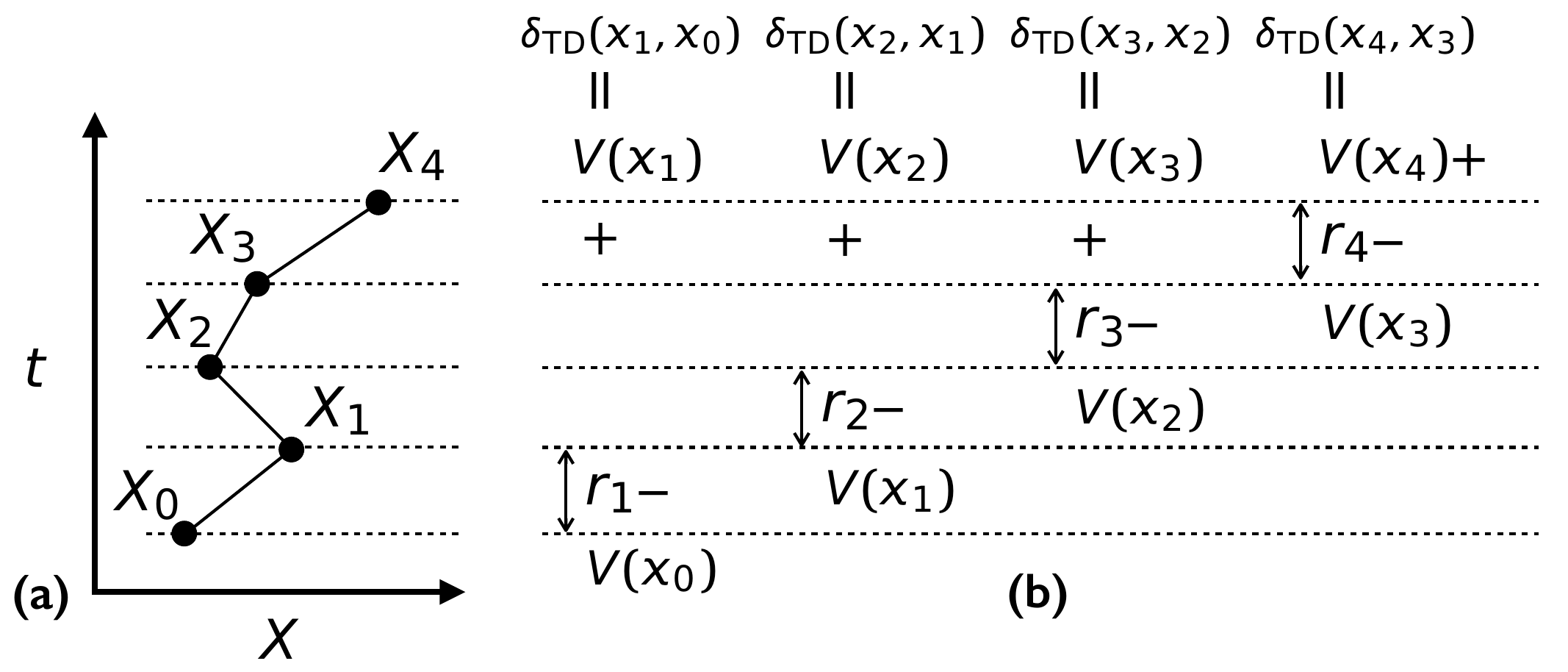}
	\end{center}
	\caption{\label{actor_critic_sketch} Sketch of the information used in updates based on 1-step actor-critic. (a) A simple sketch of an excursion, with space along the $x$ axis and time $t$ along the $y$ axis. (b) The information used to update the transitions originating from times $t=0,...,3$: the rewards $r_t$ following each state $x_t$, combined with the value $V_\psi(x_{t+1})$ of the following state $x_{t+1}$, then contrasted with the value of the prior state $V_\psi(x_t)$.}
\end{figure}

To turn this discussion into an algorithm, as before we sample some number of trajectories and then construct estimates of equations \eref{td_dynamical_gradient} and \eref{td_value_gradient}: for a single trajectory $\omega_0^T$ with temporal differences $\delta_\mathrm{TD}(x_t,x_{t-1},t)$ associated to transitions from $x_{t-1}$ to $x_t$ at time $t$, we have
\begin{eqnarray}\label{empirical_actor_critic_value}
\nabla_\psi L_V(\psi)
\approx-\sum_{t=1}^{T}\delta_\mathrm{TD}(x_t,x_{t-1},t)\nabla_\psi V_\psi(x_{t-1},t-1),
\end{eqnarray}
and
\begin{eqnarray}\label{empirical_actor_critic_policy_loss}
\nabla_\theta D_{KL}(P_\theta|P_W)
&\approx-\sum_{t=1}^{T}\delta_\mathrm{TD}(x_t,x_{t-1},t)\nabla_\theta\ln P_\theta(x_{t}|x_{t-1},t).
\end{eqnarray}
Intuitively, these updates follow exactly the discussion at the beginning of this section: along each trajectory, the value of each state is contrasted with the value of the state following it plus the reward received in between, cf.\ figure \ref{actor_critic_sketch}(b).
If the value of the resulting state combined with the reward is greater than the prior state, a contribution is added to the update which aims to increase the probability of this transition, along with the value of the prior state; the converse statements hold if the comparison is less.
For each trajectory, these contributions are then averaged in an attempt to respect all the corresponding directions.

Actor critic algorithms were among some of the earliest considered for reinforcement learning, recently returning to favour due to their ease of application to continuous state spaces, improved theoretical convergence properties over purely value focused approaches, and speed compared with purely return based policy gradient methods.
The algorithm \ref{actor_critic} presented here is closely related to the recently proposed soft actor-critic algorithm of RL \cite{Haarnoja2018}, with the key difference being the use of an initial dynamics which is targeted, rather than simply maximising entropy.

\begin{algorithm}[h]
	\caption{KL regularized actor-critic}\label{actor_critic}
	\begin{algorithmic}[1]
		\State \textbf{inputs} dynamical approximation $P_\theta(x_t,x_{t-1},t)$, value approximation $V_\psi(x_t,t)$
		\State \textbf{parameters} learning rates $\alpha^\theta_n$, $\alpha^\psi_n$; total updates $N$
		\State \textbf{initialize} choose initial weights $\theta$ and $\psi$, define iteration variables $n$ and $t$, total errors $\delta_P$, $\delta_V$, individual error $\delta$
		\State $n\gets0$
		\Repeat
		\State Generate a trajectory $\omega_0^T$ according to the dynamics given by $P_\theta(x_t,x_{t-1},t)$, with rewards $r(x_t,x_{t-1},t)$ after each state $x_{t-1}$.
		\State $t\gets0$
		\State $\delta_P\gets0$
		\State $\delta_V\gets0$
		\Repeat
		\State $\delta\gets V_\psi(x_{t+1},t+1)+r(x_{t+1},x_{t},t+1)-V_\psi(x_{t},t)$
		\State $\delta_P\gets\delta_P+\delta\nabla_\theta\ln P_\theta(x_{t+1}|x_{t},t+1)$
		\State $\delta_V\gets\delta_V+\delta\nabla_\psi V_\psi(x_t,t)$
		\State $t\gets t+1$
		\Until{$t=T$}
		\State $\theta\gets\theta+\alpha^\theta_n\delta_P$
		\State $\psi\gets\psi+\alpha^\psi_n\delta_V$
		\State $n\gets n+1$
		\Until{$n=N$}
	\end{algorithmic}
\end{algorithm}

In actor-critic algorithms a poor value approximation will clearly lead to poor or even negative changes to the dynamics.
One way to address this is by choosing learning rates in such algorithms tuned such that the value function learns faster than the dynamics, in the hope that it always provides a good approximation to the true value function for the current dynamics, and thus a good way of estimating the gradient.
So that the value approximation is relatively accurate when updates to the dynamics begin, it may also be good to have a period where only the values are updated for a fixed initial dynamics, such as the original one.
Even under these ideal conditions, actor-critic algorithms do not converge to the weights corresponding to local minima of the original loss function \eref{monte_carlo_dynamics_loss}, but have been shown to end up in a neighbourhood of such minima with high probability for linear function approximations \cite{Bhatnagar2009}.

This unavoidable inaccuracy is a result of the natural bias away from the true gradient introduced by using approximate temporal difference errors.
In many RL algorithms, this bias, causing eventual inaccuracy in the final result, is seen as the cost of the substantial reduction in the variance of gradient estimates they produce, allowing for significant improvements in convergence rates.

\subsection{Finite horizon example: random walk excursions}\label{finite-horizon}
We finish this section with a simple example of these techniques in practice, studying the excursion problem outlined in section \ref{sec:conditioned-dynamics}.
While the aim is to generate trajectories for the conditioned ensemble with weights $W(x_T,x_{T-1},T)=\delta(x_T)$ and $W(x_t,x_{t-1},t)=H(x_t)$ for $0<t<T$, due to the zero weight given to some trajectories, we must use a softened condition given by Eq. \eref{softened-constraint} and \eref{soft-excursion-distance} as a target ensemble to optimize sampling for.
This is an exactly solvable problem in the conditioned case, as outlined in \hyperref[exact-excursions]{Appendix A}, using a gauge transformation based approach which can in principle also be used calculate the exact optimal dynamics numerically for this simple softened problem.
For evaluating how well we are targeting the softened ensemble, we use this same gauge transformation technique to numerically estimate the maximum return as outlined in appendix \hyperref[maximum-returns]{Appendix B}.
We test all three algorithms currently discussed: Monte Carlo returns (MCR) shown in Alg. \ref{monte_carlo_reinforce}, Monte Carlo with a value baseline (MCVB) as in Alg. \ref{monte_carlo_reinforce_value_baseline}, and actor-critic (AC) as outlined in Alg. \ref{actor_critic}.

For simplicity we start by testing them in a simple ``tabular'' setting: that is, we associate a single weight $\theta(x,t)$ to each states transitions, and another single weight $\psi(x,t)$ to each states value for the algorithms which use them.
The transition up is then given by this weight in terms of a sigmoid
\begin{eqnarray}
	P_\theta(x+1|x,t)=\sigma(\theta(x,t))=\frac{e^{\theta(x,t)}}{e^{\theta(x,t)}+1},
\end{eqnarray}
with the probability of transition down then fixed by normalization.
The values are simply given by $V_\psi(x,t)=\psi(x,t)$.
To perform gradient descent, we need the gradients of these with respect to the weights, simply given by 
\begin{eqnarray}
	\frac{\partial\ln P_\theta(x\pm1|x,t)}{\partial\theta(x',t')}&=\pm\delta_{xx'}\delta_{tt'} P_\theta(x\mp1|x,t),
\end{eqnarray}
and
\begin{eqnarray}
	\frac{\partial V_\psi(x\pm1|x,t)}{\partial\psi(x',t')}&=\delta_{xx'}\delta_{tt'}.
\end{eqnarray}
Note that since each state has an independent weight, as signified by the Kronecker deltas, we can simply update each of these weights independently rather than storing the whole vector of updates.

\begin{figure}
	\begin{center}
		\includegraphics[width=1\linewidth]{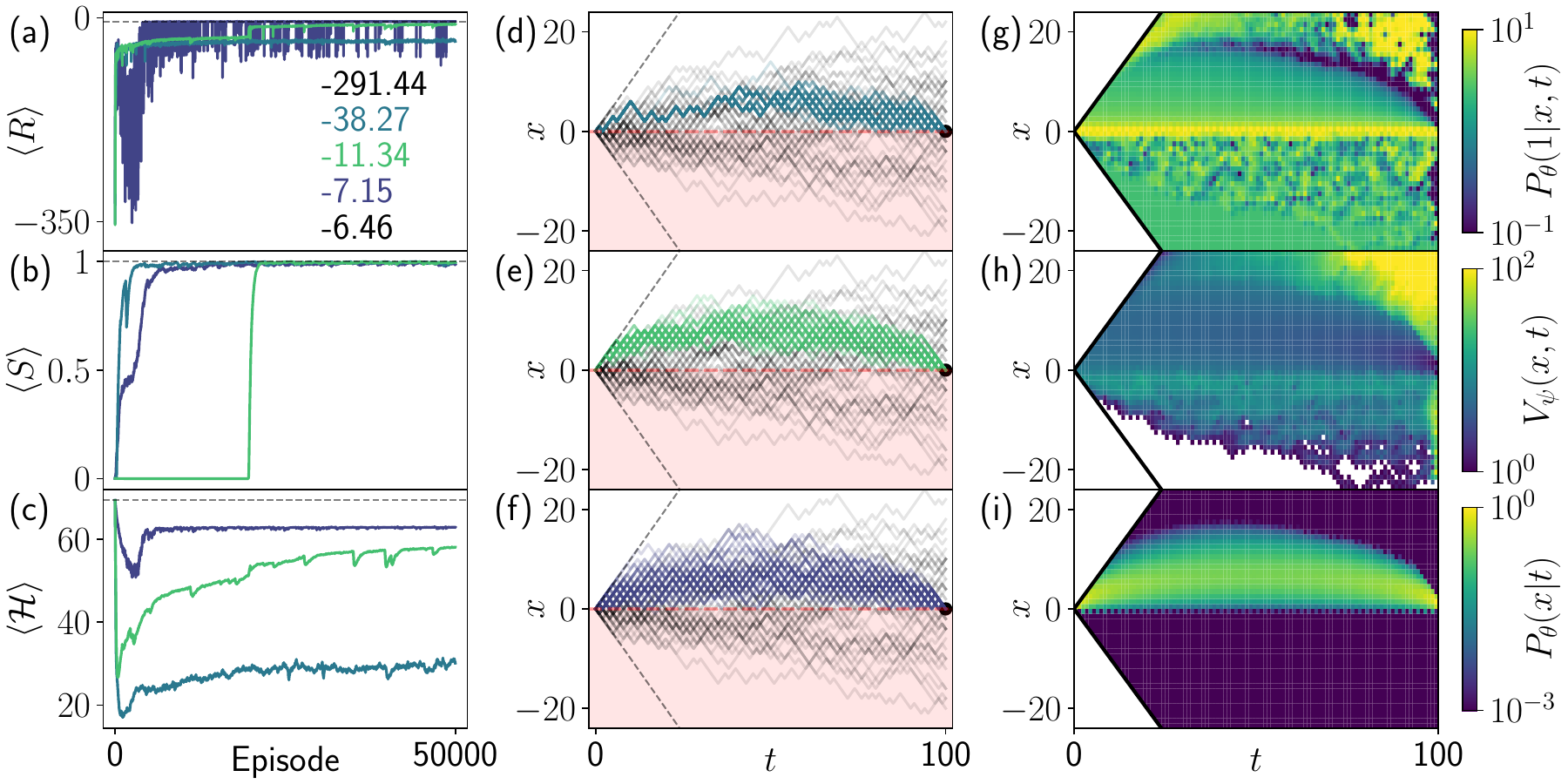}
	\end{center}
	\caption{\label{tabular_excursions} \textbf{Tabular excursions}. In these plots: AC is indicated by dark purple with and trained with $\alpha_\theta=0.15$ and $\alpha_\psi=0.3$; MCR by a lighter blue and trained with $\alpha_\theta=0.05$; and MCVB by a light green and trained with $\alpha_\theta=0.05$ and $\alpha_\psi=0.3$. The trajectory length is $T=100$ and the parameters of the softened constraint \eref{soft-excursion-distance} are $s=7$ and $a=5/7$. (a) Running averages of the returns received per episode during the learning process. The numbers indicate the initial return (top), final returns for MCR, MCVB and AC (2nd to 4th, colored) and optimal (bottom), with optimal shown by the dashed line. (b) Running averages of the probability of successfully generating an excursion. (c) Running averages of the entropy of the trajectory ensemble, with maximum $T\ln2$. (d-f) Sample trajectories generated using the final dynamics achieved for MCR, MCVB and AC (top to bottom). (g) The probability of going up at each position and time $(x,t)$ for the AC result, indicated by $P_\theta(1|x,t)=P_\theta(x+1|x,t)$ for compactness. (h) The value of each state learnt while training the dynamics using AC. (i) The probability of being in each state at each time for the final dynamics trained using AC, with normalization along each time-slice. Plots (g-i) have been interpolated over the sites which are not visited (even position, odd time, vice-versa) for visual clarity.}
\end{figure}

For evaluation of the dynamics during training, we calculate running averages of three quantities: the expected return, $\left\langle R\right\rangle_{P_\theta}$; the success rate, i.e.\ the probability of generating an excursion
\begin{eqnarray}
	\left\langle S\right\rangle=\left\langle \delta(x_T)\prod_{t=1}^{T-1}H(x_t)\right\rangle_{P_\theta},
\end{eqnarray}
which is simply the expected weighting of the conditioned ensemble; and the entropy of the trajectory ensemble
\begin{eqnarray}
	\left\langle\mathcal{H}\right\rangle=-\left\langle\ln P_\theta(\omega_0^T)\right\rangle_{P_\theta},
\end{eqnarray}
which in this case is a direct measure of the KL divergence between the optimized dynamics and the original dynamics, since $\left\langle\mathcal{H}\right\rangle=T\ln2-D_\mathrm{KL}(P_\theta|P)$.
These running averages are calculated using a learning rate and the quantities sampled from each episode: i.e.\ given a sample $O_i$ of one of the three observables from episode $i$, we update our average as $\left\langle O \right\rangle_i = \left\langle O \right\rangle_{i-1} + \alpha_O(O_i - \left\langle O \right\rangle_{i-1})$.
Observable learning rates are chosen as $\alpha_R=0.1$, $\alpha_S=0.003$ and $\alpha_\mathcal{H}=0.01$ for all three algorithms.

Results for these three quantities calculated during the learning process for excursions of length $T=100$ are shown in figure \ref{tabular_excursions}(a-c), with AC performing best on all three metrics.
In particular, we note that the AC is generally more stable, as it is less likely to get stuck in areas where the gradient of the dynamics is small, i.e.\ for large values of the potential $\theta(x,t)$.
The MC methods are vulnerable to this since they use full returns: initially, these returns may be extremely negative, particularly for earlier states if a trajectory spends a significant amount of time below $0$, causing a sudden jump to a very large value of the potential.
This can cause the dynamics to become almost deterministic for a long time (cf.\ the beginning of the samples in Fig. \ref{tabular_excursions}(d)); alternatively, the dynamics may get stuck taking incorrect actions such as going below zero for a long time, e.g. causing the initial low success rate for the MCVB training run in Fig. \ref{tabular_excursions}(b).

The slow propagation of information about the reward structure under AC training, one transition back at a time, suppresses these large negative returns early on, causing a greater emphasis on maintaining a high entropy (low KL divergence to the initial dynamics).
On the other hand, in this case the MC methods can achieve a higher return earlier by emphasising successfully generating excursions, but struggle to later optimize the entropy, due to the high variance in futures after each transition.

Plots in figure \ref{tabular_excursions}(g-h) show the upward transition probability, state values and occupation probabilities resulting from the AC training run.
The upward probabilities have the expected structure: going upwards from zero, they start at unity probability, reducing to $50-50$ along the most commonly visited set of states, and further reducing to 0 as the edge of the backwards lightcone from $x=0$, $t=0$ is reached.
After $t\sim50$, transitions upwards are suppressed earlier than the edge of the backwards lightcone, due to the rapidly reducing trajectory entropy that would result from taking further steps upwards.
The occupation probability, normalized along each time-slice, rises away from these boundaries, peaking at around $x\sim\sqrt{100}$.

Overall, for this example we can see that the resulting increase in the speed of learning more than justifies the theoretical bias induced in the final results by the various steps involved in developing these algorithms, producing results of sufficient accuracy much more quickly.

\subsection{Connection to regularized and maximum-entropy reinforcement learning}\label{sec:connect-reg-RL}
We now briefly discuss the relationship between the approach presented here and that of maximum-entropy RL \cite{Neu2017,Geist2019,Haarnoja2017,Haarnoja2018,Levine2018}.
In particular, first consider the ``deterministic'' RL case, translating from our Markov chains to an MDP by associating each transition to an action, identifying the dynamics with the RL agents policy.
Training with maximum-entropy RL is identical to training with our KL regularized algorithms, provided we choose the original dynamics to be that of the maximum-entropy trajectory ensemble, in which every trajectory has the same probability regardless of length, and the weighting is that given by biasing with respect to the reward function.

In the ``stochastic'' case, the connection is less clear.
Viewing our Markov chain as having a state space which consists of state-action pairs, and decomposing the dynamics into policy and environment components, it may be suspected that maximum-entropy RL can be recovered by choosing the original dynamics to be the one generated by a policy which produces the maximum-entropy trajectory ensemble, up to its ability to control the transitions around the environment.
However, this turns out not to be the case: such a policy would necessarily take into account the entropy of the environment resulting from each action, something which standard maximum-entropy RL does not take into account, as this would require incorporating knowledge of the environment probabilities.
Maximum-entropy RL in this case is recovered by choosing the original trajectory probabilities to consist of only the contributions of the environment, to each trajectory, normalized as required: it is not immediately clear that this ensemble itself decomposes into a Markovian structure.
This distinction may suggest a novel model-based maximum-entropy RL algorithm, in which a known or learnt model is used to further try to maximize the entropy of the trajectory ensemble over considering the policy entropy alone.

\section{A universe of algorithms: reviewing variations found in reinforcement learning}\label{sec:review}
The optimization of the KL divergence can be further manipulated in a large number of ways, each corresponding to different algorithms for approximating the gradient.
While we will not give an exhaustive coverage of the possibilities presented in the RL literature, in this section we will review some key variations, translating them into the notation used in this paper.
In particular, in section \ref{neural-networks} we demonstrate how to adapt the algorithms to train neural networks, a powerful form of function approximation.
It is hoped this will give the reader an idea of the range of techniques made available by connecting the problem of efficient trajectory sampling with RL.
However, we have made later sections independent of this one: those interested in how the approach can be specialized to the long-time limit can skip this section on first reading and instead go to section \ref{LDs}.

\subsection{Mixing estimates: expected errors, $n$-step temporal differences and weighted averages}
Here we focus on two ways of modifying the actor-critic approach, capable of reducing variance without introducing significant bias: making use of the dynamics to calculate exact expectations of temporal difference errors and gradients associated to transitions for a particular state; and using the Bellman equation to look multiple steps ahead, producing a range of equally valid estimates which can then be averaged.

Firstly, rather than manipulating the value loss into the form shown in equation \eref{td_value_gradient}, we could instead use the current dynamics to calculate the expected target for each state visited along a trajectory, as suggested by equation \eref{bellman_value_loss}, resulting in
\begin{eqnarray}\label{expected_td_value_gradient}
\nabla_\psi L_V(\psi)
\approx-\left\langle\sum_{t=0}^{T-1}\delta_{\mathbb{E}\mathrm{TD}}(x_t,t)\nabla_\psi V_\psi(x_t,t)\right\rangle_{P_\theta},
\end{eqnarray}
written in terms of the expected value of the TD error
\begin{eqnarray}
\delta_{\mathbb{E}\mathrm{TD}}(x_t,t)
&=\left\langle\delta_\mathrm{TD}(x_{t+1},x_t,t)\right\rangle_{P_\theta,X_t=x_t},
\end{eqnarray}
producing updates similar to the expected SARSA algorithm \cite{vanSeijen2009}.

Unfortunately this error can not be used for the dynamical gradient, due to the dependence of the transition on the resulting state: however, we can manipulate equation \eref{td_dynamical_gradient} to arrive at
\begin{eqnarray}\label{expected_td_dynamical_gradient}
\fl \nabla_\theta D_{KL}(P_\theta|P_W)
&\approx-\left\langle\sum_{t=1}^T\left\langle\delta_\mathrm{TD}(x_t,x_{t-1},t)\nabla_\theta\ln P_\theta(x_{t}|x_{t-1},t)\right\rangle_{P_\theta,X_{t-1}=x_{t-1}}\right\rangle_{P_\theta},
\end{eqnarray}
where for states sampled along each trajectory we calculate the expected product of the TD error and the gradient of the corresponding transition.
This possibility has recently been studied indepth in the RL literature, named variously expected policy gradients and mean actor critic \cite{Sutton2000a,Allen2017,Ciosek2018}.

In contrast to updates based on equation \eref{td_value_gradient} and \eref{td_dynamical_gradient}, updates using \eref{expected_td_value_gradient} and/or \eref{expected_td_dynamical_gradient} are reasonably expected to have much lower variance than their sampled-transition counterparts, thus resulting in improved convergence without the usual accompanying increase in bias of the final result.
The pay-off is a much higher computational demand, in part due to the need to calculate the expectation and the gradients of each transition.
Another technicality is the necessity of both updates using different quantities, whereas the updates in algorithm \ref{actor_critic} are both built around the same temporal difference errors.
It is worth noting that recent work in RL has suggested the possibility of using a mixture of both updates, with the relative weighting varying over time \cite{Asis2017}.
This may be beneficial when the most likely transitions are to states for which the value is much more accurate, reducing the propagation of errors.

Secondly, we note that the inductive Bellman equation \eref{bellman_equation} for the exact value can be substituted into itself multiple times, arriving at an $n$-step equation
\begin{eqnarray}
V_{P_\theta}(x_t,t)=\left\langle R\left(\omega_{t}^{t+n}\right)+V_{P_\theta}(x_{t+n},t+n)\right\rangle_{P_\theta,X_t=x_t},
\end{eqnarray}
which inspires an approximate $n$-step temporal difference error similar to the single step errors before
\begin{eqnarray}\label{n-step_td_error}
\delta_\mathrm{TDn}(\omega_{t}^{t+n},t)=V_{\psi}(x_{t+n},t+n)+R\left(\omega_{t}^{t+n}\right)-V_{\psi}(x_t,t).
\end{eqnarray}
Similar arguments and manipulation to that done for the 1-step temporal difference estimates of the gradients leads to the pair of approximations
\begin{eqnarray}\label{n-step_td_dynamical_gradient}
\nabla_\theta D_{KL}(P_\theta|P_W)
&\approx-\left\langle\sum_{t=1}^T\delta_\mathrm{TDn}(\omega_{t-1}^{t+n},t)\nabla_\theta\ln P_\theta(x_{t}|x_{t-1},t-1)\right\rangle_{P_\theta},
\end{eqnarray}
and
\begin{eqnarray}\label{n-step_td_value_gradient}
\nabla_\psi L_V(\psi)
\approx-\left\langle\sum_{t=}^{T-1}\delta_\mathrm{TDn}(\omega_{t}^{t+n},t)\nabla_\psi V_\psi(x_t,t)\right\rangle_{P_\theta},
\end{eqnarray}
with values and rewards which would occur at or after the end of the trajectory in the above equation set to zero.

Empirical studies of algorithms based on these errors, simply replacing the temporal difference error in \ref{actor_critic} with \eref{n-step_td_error}, suggest that each problem has an optimal value of $n$: larger values result in higher variance errors, while allowing faster propagation of reward information.
Values of $n$ greater than the trajectory length recover the Monte Carlo techniques of the previous sections. Their benefit in gradient estimation on their own merits is limited, but as we will see next, they act as a building block in a more powerful estimation scheme.

While temporal difference errors, particularly 1-step errors, result in a particularly low variance for the gradient estimates, they can result in slow propagation of information about the reward structure.
A large reward occurring on average $n$ steps in the future of a particular transition, would require at least $n$ trajectories for information about that reward to propagate back to that transition, likely many more.
In contrast, were we using an $n$-step error, reward information would propagate more quickly, but result in increased variance of the errors.

A good compromise can be achieved by observing that rather than considering any single one of the possible $n$-step approximations to the gradient, we could just as justifiably consider a weighted average of them \cite{Watkins1989,Jaakkola1994}.
That is, for some distribution $P(n)$ such that
\begin{eqnarray}
\sum_{n=1}^T P(n)=1,
\end{eqnarray}
we may consider for the dynamics
\begin{eqnarray}\label{weighted_return_dynamical_gradient}
\nabla_\theta D_{KL}(P_\theta|P_W)
\approx-\left\langle\sum_{t=0}^T\delta^P_\mathrm{TD}(\omega_{t+1}^{T},x_t,t)\nabla_\theta\ln P_\theta(x_{t}|x_{t-1},t)\right\rangle_{P_\theta},
\end{eqnarray}
with the weighted error
\begin{eqnarray}
\delta_\mathrm{TD}^P(\omega_{t}^{T},t)=\sum_{n=1}^{T-t}P(n)\delta_\mathrm{TDn}(\omega_{t}^{t+n},t),
\end{eqnarray}
and a similar equation for the value loss gradient.
Special cases of the distribution defining this error provide both the Monte Carlo and temporal difference errors discussed previously, however, we can now perform updates according to an equal weighting of the Monte Carlo and 1-step errors in each trajectory, or any other distribution we choose.
Depending on this choice, we can achieve much faster propagation of information about the reward structure.
Further, we can tune the distribution to minimize both the effect of the increased variance inherent in the considering more of the future of each sampled trajectory, and the effect of inaccurate value functions replacing the future.

A common distribution chosen in an attempt to achieve a balance between the variance of longer $n$-step errors and propagation of reward information is a normalized geometric series
\begin{eqnarray}
P(n)=\frac{\lambda^{n-1}(1-\lambda)}{1-\lambda^T},
\end{eqnarray}
which allows for efficient numerical implementation to be achieved by deriving inductive equations relating this return to its value at the next time step.

For completeness, we also note that the expected TD error can be extended in an $n$-step or $\lambda$-weighted form, related to the so-called Tree-Backup algorithm in RL \cite{Precup2000}.
Studies of $n$-step or $\lambda$-weighted adaptations of mean actor critic have yet to be conducted.

\subsection{Online learning, importance sampling and eligibility traces}\label{online_offpolicy_eligibility}
In this subsection we briefly discuss a trio of related RL techniques.
First, many RL algorithms are designed to be implemented in an online manner, that is, updates may be applied after every transition, not after the end of each trajectory.
This allows for experiences during the current trajectory to be used immediately, potentially leading to faster convergence, and as we will see in the next section is essential for infinite-horizon problems where trajectories do not end, rendering Monte Carlo methods impossible.

For a simple heuristic justification of this, note we may rewrite the gradients for the 1-step TD approximations as
\begin{eqnarray}\label{online-gradient}
\nabla_\theta D_{KL}(P_\theta|P_W)
&\approx-T \, \left\langle\delta_\mathrm{TD}(x_t,x_{t-1},t)\nabla_\theta\ln P_\theta(x_{t}|x_{t-1},t)\right\rangle_{P_\theta},\\
\nabla_\psi L_V(\psi)
&\approx-T \, \left\langle\delta_{\mathrm{TD}}(x_{t},x_{t-1},t)\nabla_\psi V_\psi(x_{t-1},t-1)\right\rangle_{P_\theta},
\end{eqnarray}
where we are now viewing the expectation as sampling the triplet of a pair of consecutive states at a particular time, with time is sampled uniformly according to $1/T$.
The pair of states are sampled at that time according to the state distribution and transition probabilities of the current dynamics.
In reality, we produce correlated samples of this expectation by running trajectories, with the time of each sample being iterated along by one from the previous time.
Ignoring technicalities caused by the correlations of the samples generated, from this perspective online algorithms simply apply stochastic gradient descent at the level of individual transitions, rather than individual trajectories.

We do, however, note a subtlety in this viewpoint: by using online updates during the sampling of trajectories, the transitions leading up to the current time are not sampled according to the current dynamical weights, but instead sampled according to the weights at the moment that transition was simulated.
Thus, for the heuristic SGD perspective above to be completely valid, we would have to use an importance sampling factor to take into account the true probability of having arrived in the present state under the current dynamical weights.
In practice, the small bias this induces is tolerated, as this importance sampling factor would be difficult to implement.

Importance sampling arises more commonly in RL through off-policy methods, in which data is collected using an alternative dynamics to the one being optimized.
In this context we must take into account the alternative sampling probabilities twice: reweighting the past to account for the different likelihoods of arriving in a particular state at a particular time, and reweighting the errors themselves to account for the chance of the sampled transition occurring.
The later is easy to compensate for, while the former is in principle a complex ratio of historical probabilities.
For the values, ignoring the former is equivalent to choosing an alternative prioritization for which states to optimize with respect to.
When using the semi-gradient methods described earlier, if this shifted priority differs too substantially from the current dynamics, this can result in a lack of convergence in learning algorithms; if close enough, the dynamics will converge, but be biased further away from the ideal weights \cite{Sutton2000,Phansalkar1995}.
Since the effect in the prioritization of online learning will be minor, this later point is suggestive of the effect this will have on a learning algorithms results: while the weights would be expected to converge, perhaps faster than an offline approach, the end result may be less accurate than the best possible from offline learning.

While true stochastic gradient methods can address the lack of convergence in off-policy sampling \cite{Sutton2009,Maei2009,Maei2011,Maei2018}, they do not address the incorrect priority of states.
For the dynamics, ignoring the importance sampling ratio for the history is even more detrimental, implying we are not estimating a gradient of the loss function \eref{KL_divergence} which our main goal it is to minimize.
We should therefore handle this lack of emphasis on the correct states in order to reach optimal weights.
Off-policy policy gradient techniques are an open area of research in RL \cite{Degris2012}, however, progress has recently been made through techniques which estimate what the correct emphasis to give states \cite{Maei2018,Imani2018}.
Despite the bias this emphasis induces in principle, removing it is difficult enough that many state-of-the-art algorithms forgo doing so, accepting any potential reduction in the quality of the final result.

Online learning may be used instantaneously with 1-step errors or temporarily delayed for $n$-step errors.
The weighted $\lambda$-errors can also be approximately implemented completely online through the use of so-called eligibility traces, closely related to Malliavin weights \cite{Warren_2013}. 
These approximate the true $\lambda$-error updates, due to the continual drift of the weights away from those associated to the particular transition the $\lambda$-error is being calculated for \cite{Watkins1989,Sutton1988,Precup2000,Degris2012,Sutton2018}.
For linear function approximations this drift can be compensated efficiently, leading to very effective algorithms, however, for general non-linear functions the approximate nature of more general eligibility trace methods can in fact prevent convergence and lead to poor results \cite{Seijen2015}.
It may thus be more desirable to implement $\lambda$-errors online by first truncating them to $n$-steps, then applying delayed updates calculated iteratively for equivalent computational complexity as eligibility trace approaches, at the expense of increased memory requirements \cite{Cichosz1995,Seijen2016,Veeriah2017}.
However, as we discuss next, even taking this approach may result in instability for common non-linear function approximations.

\subsection{Using neural networks: replay buffers and target networks}\label{neural-networks}
A powerful function approximation that has found substantial use across academia and industry in recent years is that of neural networks.
Unfortunately, while powerful, training them in the straightforward manner described previously often proves to be extremely unstable.
This is a consequence of the so-called ``catastrophic interference'' that neural networks suffer from: their strong adaptability and broad representational power is accompanied by a tendency to forget all but the most recent experiences used in training them.
In supervised and unsupervised problems this causes issues in sequentially learning one problem after another, transferring a learned network to a new problem, or when the data distribution is non-stationary in some real-world applications \cite{McCloskey1989,Ratcliff1990,Kirkpatrick2017,Riemer2018}.
This can be traced back to correlations in the data samples used in training, resulting in non-IID sampling: in sequential or transfer learning, samples are correlated by the simple fact that they belong to one problem or another.
While this issue also exists in transferring learned policies and value functions between control problems, in RL, catastrophic interference can in fact occur during training on individual problems, as data is naturally correlated when sampled from trajectories using a Markovian dynamics \cite{Ghiassian2018,Nguyen2019,Lo2019}.
Often experienced most severely in online training, we even observed this phenomenon during offline training if the samples from a trajectory are strongly correlated, such as in the excursion problem of section \ref{finite-horizon}.
Further to this, RL is a highly non-stationary problem, with both the state distribution changing whenever the policy is updated, and the targets used in estimating the gradient changing whenever the value function is updated.

As a straightforward demonstration on the simple excursion problem discussed above, we chose to generate batches of $64$ trajectories between each update, constructing estimates of both the policy and value gradients using the actor critic algorithm \ref{actor_critic}, averaging the temporal difference errors for transitions present in the batch of trajectories.
We used neural networks with input tuples of $(x,t)$, processed through two 64 neuron hidden layers and one 32 neuron hidden layer for both the policy and value function, with the first two layers followed by a ReLu activation function: for the value function the final layer was linear, while for the policy this was followed by a sigmoid to return a probability between $0$ and $1$ for transitioning up.
Learning rates for both networks were chosen to be a constant $\alpha^\theta=\alpha^\psi=0.0004$.
For the weighting, c.f.~\eref{softened-constraint} and \eref{soft-excursion-distance}, we used $sb=-50$ reward for transitions to a negative position; for transitions to the final time state, the exponent is modified to a linear dependence on the final position with $s=500$, $W(x_T,T)=\exp(-500|x_{T}|)$.

\begin{figure}
	\begin{center}
		\includegraphics[width=0.666\linewidth]{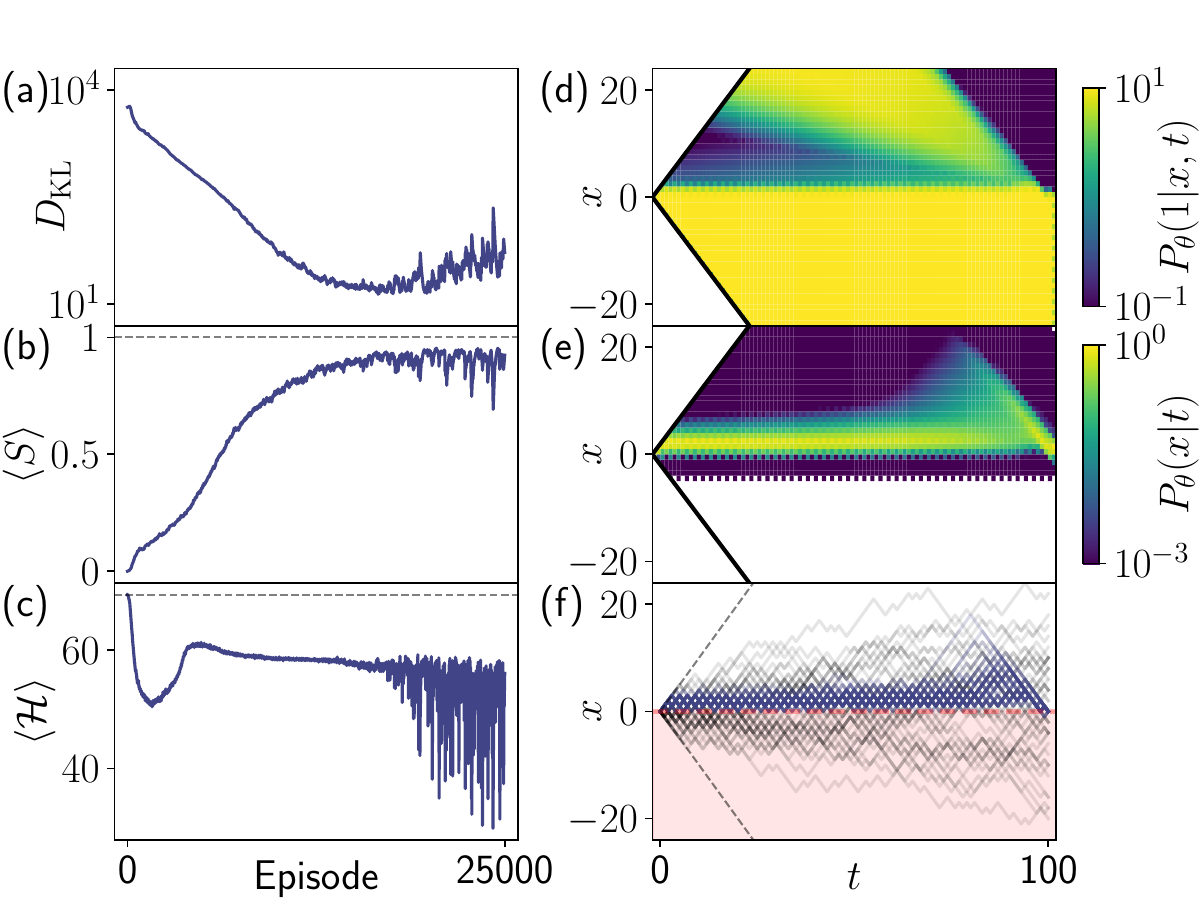}
	\end{center}
	\caption{\label{neural_excursions} \textbf{Neural-network excursions}. Here the trajectory length is $T=100$. (a) An estimate of the KL divergence using a running average over returns from the batch of trajectories generated between each update. (b) Running average over each batches probability of successfully generating an excursion. (c) Running average over each batches trajectory entropy. (d) The probability of going up at each position and time $(x,t)$ for the final result, indicated by $P_\theta(1|x,t)=P_\theta(x+1|x,t)$ for compactness. (e) The probability of being in each state at each time for the final dynamics, with normalization along each time-slice. Plots (d,e) have been interpolated over the sites which are not visited (even position, odd time, vice-versa) for visual clarity. (f) Sample trajectories generated using the final dynamics achieved. }
\end{figure}

Results of this optimisation are shown in figure \ref{neural_excursions}.
Analogous to figure \ref{tabular_excursions}, \ref{neural_excursions}(a-c) show running averages of the KL divergence, success rate and trajectory ensemble entropy during the learning process.
While the KL divergence remains much larger than the equivalent for the tabular approach, this is largely due to the magnitude of the weight exponents used: for example, note the initial KL divergence is on the order of $10^4$, in comparison to order $10^2$ for the tabular results, despite beginning at the same maximum entropy dynamics.
Although a significant improvement over the original dynamics, the success rate and entropy do not quite achieve the levels seen in the tabular approach, owing to the difficulty in overcoming the instability mentioned above in order to optimize neural networks to a high degree of accuracy.
The entropy in particular is lower than desired: the final up-transition probabilities in figure \ref{neural_excursions}(d) show a significant region when the network has learnt to go up at higher values of the position, until the upper edge of the backwards lightcone from the target is reached.
This is bordered below by a region where transitions down are almost certain, likely a result of the current dynamics closer to $x=0$ being more entropic, and thus rewarding, compared to the higher position dynamics which simply goes up to the lightcone edge before going down.
The resulting state distribution \ref{neural_excursions}(e) (with sample trajectories demonstrated in \ref{neural_excursions}(f)) is far more focused around $x=0$ than we would hope for, as seen in \ref{tabular_excursions}(i).
These issues likely stem from the large exponents used for the weights, dominating the contribution of the entropy in the KL divergence.
This makes it difficult for the learning algorithms to ``see'' the entropy past the potentially large negative weight contributions, making optimization of the entropy a slow process which can't be achieved before the training becomes unstable.

This instability in training neural networks with RL algorithms starts to become pronounced at longer training times, as seen by the noise present at the end of all three learning curves and the increasing value of the KL divergence, even while averaging over a large number of trajectories.
More generally, in order to train a neural network, a variety of stabilizing techniques are often used, aimed at suppressing correlations between training samples \cite{Mnih2013,Mnih2015,Lillicrap2015,Wang2016,Haarnoja2018}.
Typically, two main adaptations are used.

For the non-stationarity of the values used in bootstrapping estimates of the gradients, a third ``target'' network is introduced: this is either periodically updated to the current weights of the value network, remaining fixed while the value network is updated in between \cite{Mnih2013,Mnih2015,Lillicrap2015}, or slowly updated toward the current weights after each update of the value network using an exponential average \cite{Haarnoja2018}.
However, the instability caused by these moving targets is largely a result of the semi-gradient approximation we made, and can alternatively be addressed instead by using the gradient TD methods \cite{Sutton2009,Maei2009,Maei2011,Maei2018} mentioned in section \ref{sec:actor-critic}, which take into account the change in the target by considering its derivative.

Meanwhile, both the non-stationarity of the state-distribution and the correlation of trajectory-based sampling are partially addressed by the introduction of experience replay \cite{Lin1992,Mnih2013,Mnih2015,Wang2016,Haarnoja2018}: for example, a recent history of experienced transitions are stored in a replay buffer, from which we sample a random set of transitions for use in estimating the gradient.
This sampling from the replay buffer reduces correlations between the samples used, as they are no longer sampled sequentially from a trajectory, and slows the change in the state distribution, at the expense of biasing the updates away from their true values for the current weights.

As an example, we now cover the use of experience replay in 1-step AC algorithms in more detail.
In this case, the basic information we store in the buffer $\mathcal{D}$ are individual transitions $(x,t,x')$.
Rewards are then recalculated using the current dynamics whenever the transition is resampled from the buffer.
The bias introduced by experience replay is a result of the differing probabilities of sampling state-state pairs corresponding to each transition, between the distribution of the current dynamics and the distribution of stored in the replay buffer.
These probabilities can be decomposed into two parts: the probability of being in the state pre-transition, and the probability of that transition occurring.
We can address the later of these easily. 
If we additionally store the probability $\mu$ of each transition at the time it was originally generated, we can multiply its contribution to the gradient when resampled by an importance sampling factor $P_\theta(x'|x,t)/\mu$, removing the resulting bias.
The former of these is much more complicated to address, and as such the bias it causes is often accepted in pay off for the benefits of using a replay buffer.
However, there exist various techniques which can be used to emphasise states more appropriately in the replay buffer \cite{Maei2018,Imani2018}.
Given the correction for the transition bias, a gradient estimate is then constructed using a set of $N$ samples ${(x_i,t_i,x_i,\mu_i)}$ randomly taken from the buffer, using 
\begin{eqnarray}
\nabla_\psi L_V(\psi)
\approx-\sum_{i=1}^{N}\frac{P_\theta(x_i|x_i,t_i)}{\mu_i}\delta_\mathrm{TD}(x_i,x_i,t_i)\nabla_\psi V_\psi(x_i,t_i-1),
\end{eqnarray}
and
\begin{eqnarray}
\nabla_\theta D_{KL}(P_\theta|P_W)
&\approx-\sum_{t=0}^{T}\frac{P_\theta(x_i|x_i,t_i)}{\mu_i}\delta_\mathrm{TD}(x_i,x_i,t_i)\nabla_\theta\ln P_\theta(x_i|x_i,t_i),
\end{eqnarray}
to update the weights.

Despite the limitations of our demonstration in comparison with our earlier tabular results, we believe these could be resolved by better tuning of algorithm parameters and use of the techniques mentioned above.
Regardless, it is likely that to apply these techniques to more complex systems neural networks will be extremely useful if not essential.
For simple or complex problems, even if the optimal dynamics can not be reached, the resulting dynamics could be combined with techniques such at TPS to efficiently gather accurate statistics of the rare trajectories of interest.

Finally, we mention that while eligibility traces are powerful when used with tabular methods or linear approximations, the lack of ability to train neural networks using incremental data hinders their use.
To this end, recent work has been done considering truncated $\lambda$ returns \cite{Seijen2016,Veeriah2017}, and their reconciliation with experience replay \cite{Daley2018}.

\subsection{Further variations}\label{finite-summary}
We briefly mention a variety of other possibilities from the RL literature to approach optimizing such problems:
\begin{itemize}
	\item All algorithms described above are based on stochastic gradient descent, a commonly used line-search gradient method.
	Recently, RL algorithms have been developed based on natural gradients \cite{Kakade2001a,Peter2003,Bagnell2003,Bhatnagar2009,Thomas2014}, where the updates are modified to respect that changing the parametrization of the dynamics, while leaving the manifold of possible dynamics invariant, should leave the gradient updates invariant.
	These are closely related to recent applications of trust-region based gradient methods to RL \cite{Schulman2015a,Schulman2017,Nachum2017a,Wu2017}, where the learning rates for updates are tamed in order to try and ensure updates do not overshoot and cause a negative change to the dynamics.
	
	\item As value functions are learnt from early experiences, transitions towards states that are currently estimated to be higher value will be increased, even if these states are in reality suboptimal, a problem referred to as maximization bias.
	A common solution to this is the use of double learning, where two value functions are learnt \cite{Hasselt2010,Fujimoto2018,Haarnoja2018}.
	For each state visited, the value function which produces the lower estimate is then used in estimates of the dynamics gradients.
	
	\item When the action space is continuous, the MDP problem can be rephrased as learning a function approximation which generates an action, with inputs as the state and some random noise \cite{Silver2014,Lillicrap2015,Haarnoja2018}.
	This leads to policy gradient estimate which takes into account how the target value changes when when the action parametrization changes, resulting in a lower variance estimate.
	This will be directly relevant to rare trajectory problems with continuous state spaces and an uncountable number of transitions, and is closely related to current optimal force learning approaches in diffusive problems \cite{Das2019}.
	
\end{itemize}

An alternative but closely related adaptive approach is based on gauge transformations \cite{Garrahan2016}.
While there are simpler derivations, see \hyperref[exact-excursions]{Appendix A}, to see this connection note we may rewrite equation \eref{monte_carlo_dynamics_loss} as
\begin{eqnarray}
\fl D_{KL}(P_\theta|P_W)&=\sum_{t=0}^T\sum_{\omega_0^{t-1}}P_\theta(\omega_0^{t-1})
D_{KL}\left(P_\theta(-|x_{t-1},t)\left|\frac{W(-,x_{t-1},t)P(-|x_{t-1})g(-,t)}{g(x_{t-1},t-1)}\right.\right)\nonumber\\
\end{eqnarray}
where
\begin{eqnarray}\label{1-step_gauge}
g(x_t,t)=\mathbb{E}_{x_{t+1}\sim P}\left[W(x_{t+1},x_{t},t+1)g(x_{t+1},t+1)\right],
\end{eqnarray}
with $g(x_T,T)=1$ is the inductive equation defining the gauge transformation $g$, with expectation taken over the original dynamics.
Since minimizing each of these KL-divergences individually provides the exact solution, the optimal dynamics is given by the correct gauge transformation, and an alternative approach may be to approximate this gauge transformation directly.
This approach has a long history in the mathematical literature \cite{Borkar2002,Borkar2003,Basu2008,Borkar2010}, and as exact solutions to some MDPs with deterministic environments \cite{Todorov2009}.
Further, this has recently been adapted to diffusion processes \cite{Ferre2018}.
It has also been discussed recently in the context of understanding reinforcement learning from a statistical physics perspective \cite{Rahme2019}.
From the RL perspective, these algorithms are all based on 1-step temporal difference methods, where equation \eref{1-step_gauge} is viewed as a non-linear Bellman equation \cite{Hasselt2019}.
This approach could in future be developed into a broader set of RL algorithms which have more in common with the value-function based methods of RL, as opposed to the policy-gradient-like methods presented in this work.

\section{Long time dynamics, large deviations and discounting}
\label{LDs}

In many problems of relevance to physical sciences we are interested in the behaviour at long times, such that the system is in its stationary state, be it equilibrium (as in a system in contact with a thermal bath) or not (as in driven systems). Such situations where dynamics is time-homogeneous and the relevant times exceed those set by all relaxation rates, pertain to the regime of dynamical large deviations \cite{Touchette2009,Lecomte2007b,Garrahan2018,Jack2019}, an approach akin to equilibrium statistical mechanics for quantifying the statistical properties of long-time dynamics. For this kind of problem we can specialize our methods above to allow for solutions using genuine, infinitely long trajectories.

To consider these problems, for simplicity we restrict to cases where the original dynamics is time-independent, although the approach may be adapted to periodic dynamics. We can then consider the stationary state of some parameterized dynamics $P_\theta(x_t|x_{t-1})$, a probability distribution $P_\theta^\mathrm{ss}(x)$ such that
\begin{eqnarray}
	P_\theta^\mathrm{ss}(x)=\sum_{x'}P_\theta(x|x')P_\theta^\mathrm{ss}(x').
\end{eqnarray}
For clarity, we will focus on systems with ergodic dynamics.
Put simply, this means that for any pair of states, there exists a sequence of transitions which leads from either one to the other.
For us, this means that there is a unique stationary state.

A common approach to studying such models is to consider long but finite trajectories, then use a method such as TPS to sample the reweighted ensemble.
While we could take a similar approach using our adaptively learnt dynamics, either with or without TPS, the trajectory lengths may need to be extremely long to achieve accurate results, and for a generic problem the length required is unknown.
It may instead be desirable to directly study the infinite-horizon case, removing fears of incorrect results caused by finite-time effects.
However, as it stands there are several problems with the algorithms presented ealier in section \ref{dynamical_gradients} for studying problems formulated with an infinite-horizon.
In particular, the algorithms we detailed were ``offline'', that is, they waited for trajectories to end before learning occurred: clearly in an infinite-horizon context where there is no end to a trajectory, we must necessarily use an online approach, as discussed in section \ref{online_offpolicy_eligibility}.

There is a second, more substantial issue: as currently defined, the returns, and thus the resulting values, could diverge to infinity as the trajectory continues to run.
Moreover, the value of each state would be almost identical even for sufficiently long but finite futures, as it would be dominated by the average return following states sampled from the stationary state distribution.
The origin of these issues can be attributed to the fact that we provide equal emphasis to the value of a state for transitions which occur at any time in the future: for an ergodic system in which any correlation with the current state will eventually be lost, such a definition of value ignores the eventual independence of future states and transitions on the present state being valued.

In this section we will consider a pair of adaptations which remedy this failing of the finite-time value, so that online algorithms can be developed for the infinite-horizon case.
First, we will discuss the differential returns and relative values arising from the average-return formulation of RL; second, we will introduce an approximate scheme based on discounting, which nonetheless can improve learning speed by reducing variance, at the expense of accuracy in the final result.

\subsection{Comparing rewards with the average: differential returns and values}
For RL problems involving an infinite-horizon, one choice of formulation, sometimes argued to be the correct formulation over the traditional one based on discounting \cite{Sutton2018,Naik2019}, is that of time-averaged returns \cite{Marbach2003,Schwartz1993,Bertsekas1996,Tsitsiklis1999}.
For us, this approach begins by reconsidering our loss function.
In the continuing case, under the conditions of time-independence and ergodicity we mentioned in the previous section, there is no particular special time, such as when the trajectory is initialized.
As such, the time averaged KL divergence is simply given by a steady state average of rewards on the next transition
\begin{eqnarray}\label{continuing_KL_divergence}
d_{KL}(P_\theta|P_W)
&=\lim_{T\rightarrow\infty}\frac{1}{T}D_{KL}(P_\theta|P_W)\nonumber\\
&=-\lim_{T\rightarrow\infty}\frac{1}{T}\left[\sum_{\omega_0^T}P_\theta(\omega_0^T)R(\omega_0^T)-\ln Z\right]\nonumber\\
&=-\sum_{x,x}P_\theta^\mathrm{ss}(x)P_\theta(x|x)r(x,x)+z,
\end{eqnarray}
where we have simply defined
\begin{eqnarray}\label{scgf}
z=\lim_{T\rightarrow\infty}\frac{1}{T}\ln Z,
\end{eqnarray}
and $r(x,x)$ is the time-independent reward associated to this transition
\begin{eqnarray}
r(x',x)=\ln W(x',x) -
\ln\left(\frac{P_\theta(x'|x)}{P(x'|x)}\right).
\end{eqnarray}
For later clarity, we define the time-averaged return as
\begin{eqnarray}\label{average_return}
\bar{r}_\theta=\lim_{T\rightarrow\infty}\frac{1}{T}\sum_{\omega_0^T}P_\theta(\omega_0^T)R(\omega_0^T)=z-d_{KL}(P_\theta|P_W).
\end{eqnarray}
As we will discuss further in section \ref{sec:ld-scgf}, $z$ is related to the SCGF which is often of interest in large deviation studies.
The connection between $z$ and the average reward thus means our algorithms provide an estimate of the SCGF in the process of optimizing the dynamics.

While not immediately obvious from equation \eref{continuing_KL_divergence}, the gradient of $d_{KL}(P_\theta|P_W)$ can infact be written in terms of only the gradient of $P_\theta(x'|x)$, without reference to the gradient $P_\theta^\mathrm{ss}(x)$: that this is possible essentially follows from the fact that the steady state is defined by the dynamics.
This is extremely useful numerically, as while the gradient of the stationary state may be extremely difficult to construct, the gradient of the transition probabilities is directly accessible using our approximation.
However, to see this form of the gradient of equation \eref{continuing_KL_divergence} clearly, we must first define values in this continuing setting.

In order to construct useful values for states in the continuing case, we consider returns defined relative to the average of equation \eref{average_return}: that is, we define the differential return
\begin{eqnarray}
R_D(\omega_0^T)&=R(\omega_0^T)-T\bar{r}_\theta\nonumber\\
&=\sum_{t=1}^Tr(x_t,x_{t-1})-\bar{r}_\theta.
\end{eqnarray}
We can then consider the value of a state to be the difference between the average return following that state, and the average return following a state drawn from the stationary distribution, simply given by the average of differential returns following that state
\begin{eqnarray}
V_{P_\theta}(x_0)=\lim_{T\rightarrow\infty}\left\langle R_D(\omega_0^T)\right\rangle_{P_\theta,X_0=x_0},
\end{eqnarray}
where the limit is now convergent, as seen in the next section.
In particular, we may relate these values iteratively in a Bellman equation as
\begin{eqnarray}\label{differential_bellman}
V_{P_\theta}(x')=\sum_{x}P_\theta(x|x')\left[V_{P_\theta}(x)+r(x,x')-\bar{r}_\theta\right],
\end{eqnarray}
which can be simply rearranged to give an alternative equation for our time-averaged KL divergence
\begin{eqnarray}
d_{KL}(P_\theta|P_W)=z-\sum_{x}P_\theta(x|x')\left[V_{P_\theta}(x)+r(x,x')-V_{P_\theta}(x')\right],
\end{eqnarray}
which we note holds for all $x'$.

We can now write the gradient of our loss as
\begin{eqnarray}
\fl\nabla_\theta d_{KL}(P_\theta|P_W)
=&-\sum_{x}\nabla_\theta P_\theta(x|x')\left[V_{P_\theta}(x)+r(x,x')-V_{P_\theta}(x')\right]\nonumber\\
\fl&-\sum_{x}P_\theta(x|x')\left[\nabla_\theta V_{P_\theta}(x)-\nabla_\theta V_{P_\theta}(x')\right].
\end{eqnarray}
Since this equation holds for all $x'$, we are free to average the right hand side over the stationary state
\begin{eqnarray}
\fl\nabla_\theta d_{KL}(P_\theta|P_W)
&=-\sum_{x,x'}P_\theta^\mathrm{ss}(x')\nabla_\theta P_\theta(x|x')\left[V_{P_\theta}(x)+r(x,x')-V_{P_\theta}(x')\right]\nonumber\\
\fl&\quad-\sum_{x,x'}P_\theta^\mathrm{ss}(x')P_\theta(x|x')\left[\nabla_\theta V_{P_\theta}(x)-\nabla_\theta V_{P_\theta}(x')\right]\nonumber\\
\fl&=-\sum_{x,x'}P_\theta^\mathrm{ss}(x')\nabla_\theta P_\theta(x|x')\left[V_{P_\theta}(x)+r(x,x')-V_{P_\theta}(x')\right]\nonumber\\
\fl&\quad-\sum_{x}P_\theta^\mathrm{ss}(x)\nabla_\theta V_{P_\theta}(x)+\sum_{x'}P_\theta^\mathrm{ss}(x')\nabla_\theta V_{P_\theta}(x'),
\end{eqnarray}
where by using the definition of the stationary state and the normalization of the transition probabilities, the last two terms are seen to be equal.
Rewriting the gradient using $\nabla f=f\nabla\ln f$ we arrive at a quantity that can be sampled using transitions from trajectories
\begin{eqnarray}\label{differential_dynamics_loss}
\fl\nabla_\theta d_{KL}(P_\theta|P_W)
=-\sum_{x,x'}P_\theta(x|x')P_\theta^\mathrm{ss}(x')\left[V_{P_\theta}(x)+r(x,x')-V_{P_\theta}(x')\right]\nabla_\theta \ln P_\theta(x|x'),\nonumber\\
\end{eqnarray}
which depends only on the gradient of the transitions.

This derivation has naturally left us with a baseline of the exact value function: the second value function term in this equation could be removed by conducting the sum over $x$.
Indeed, if we introduce a baseline of $\bar{r}_\theta$ for all states, then the term in the bracket is the temporal difference error resulting from rearranging equation \eref{differential_bellman}.
The gradient is then already in the form of those considered for the actor-critic algorithms, with the critic in this case still providing the perfect values of each state.

To arrive at a functioning algorithm, we must again introduce a learnt critic.
We do this as before: we target the true values $V_{P_\theta}$ with an approximation $V_\psi$, with a loss function given by the error in the Bellman equation \eref{differential_bellman} averaged over the stationary state
\begin{eqnarray}
\fl L_V(\psi')=\sum_{x'}P_\theta^\mathrm{ss}(x')\frac{1}{2}\left[\sum_xP_\theta(x|x')\left[V_{\psi}(x)+r(x,x')\right]-\bar{r}_\theta-V_{\psi'}(x')\right]^2,
\end{eqnarray}
noting that the target from the right of the Bellman equation is fixed to the current weights $\psi$, taking a semi-gradient approach.
The gradient evaluated at the current weights $\psi$ is then
\begin{eqnarray}\label{differential_value_loss}
\fl\nabla_\psi L_V(\psi)\approx-\sum_{x,x'}P_\theta(x|x')P_\theta^\mathrm{ss}(x')\left[V_\psi(x)+r(x,x')-\bar{r}_\theta-V_\psi(x')\right]\nabla_\psi V_\psi(x'),
\end{eqnarray}
the same as equation \eref{td_value_gradient} up to the negation of the average off of the reward at each transition.

An added complexity comes from the presence of this average return, as both gradient estimates still assume we know the average exactly, which will almost certainly not be true.
We must therefore also estimate this average return during our optimization.
To do this, we could simply use the stochastic approximation with the rewards sampled over time.
Were the dynamics fixed, this would eventually converge to the correct value;
for dynamics that are optimized over time, this will continually chase the current value of the average, similar to how the weights of the value function chase the optimal weights for the current dynamics.
However, we can speed up convergence, admittedly to a less accurate result, by using the the temporal difference error.

More precisely, we can rearrange the Bellman equation and average to get
\begin{eqnarray}
\bar{r}_\theta
&=\sum_{x,x'}P_\theta^\mathrm{ss}(x')P_\theta(x|x')\left[V_{P_\theta}(x)+r(x,x')-V_{P_\theta}(x')\right],
\end{eqnarray}
which we can sample directly by running trajectories with the current dynamics.
Replacing the exact values with our current estimates, we can then update our estimate of the average $\bar{r}_n$ every time a transition occurs, e.g. from $x'$ to $x$, as
\begin{eqnarray}\label{average_td_update}
\bar{r}_{n+1}=\bar{r}_n+\alpha_n\left[V_\psi(x)+r(x,x')-\bar{r}_n-V_\psi(x')\right].
\end{eqnarray}
To make a functioning algorithm, we then replace $\bar{r}_\theta$ in the above gradient estimates for the dynamical and value approximations with our current estimate $\bar{r}_n$.

With the equations \eref{differential_dynamics_loss}, \eref{differential_value_loss} and \eref{average_td_update} in these forms, the updates for all three components -- the dynamical weights $\theta$, the value weights $\psi$, and the approximate $\bar{r}$ -- can be estimated using the same temporal difference at each step, namely
\begin{eqnarray}
\delta_\mathrm{DTD}(x',x)=V_\psi(x)+r(x,x')-\bar{r}_n-V_\psi(x'),
\end{eqnarray}
where the subscript DTD stands for ``differential temporal difference''.
The online algorithm \ref{differential_actor_critic} based on this average construction, updating the two weights and the average at every transition, is stated below.
Removing the components related to the average in this algorithm will provide an online algorithm which could easily be applied in the finite-horizon case.

As discussed in section \ref{online_offpolicy_eligibility}, online algorithms introduce two issues.
First, with the evolving weights, we almost certainly are not sampling the current stationary state of the dynamics: however, if the dynamics evolves slowly enough, the sampling is likely very similar, and certainly close enough to be confident of convergence.
Second, the samples we get are not uncorrelated, like we would ideally have in constructing an empirical mean.
For simple function approximations this is not an issue if correlations between samples decay quickly enough, however, as mentioned in section \ref{neural-networks}, for more powerful function approximations such as neural networks this can cause instability.

\begin{algorithm}[h]
	\caption{KL regularized differential actor-critic}\label{differential_actor_critic}
	\begin{algorithmic}[1]
		\State \textbf{inputs} dynamical approximation $P_\theta(x,x')$, value approximation $V_\psi(x)$
		\State \textbf{parameters} learning rates $\alpha^\theta_n$, $\alpha^\psi_n$, $\alpha^R_n$; total updates $N$
		\State \textbf{initialize} choose initial weights $\theta$ and $\psi$, initial average $\bar{r}$, define iteration variable $n$, individual error $\delta$
		\State $n\gets0$
		\Repeat
		\State Generate a transition from $x'$ to $x=\{x,F(x,x')\}$ according to the dynamics given by $P_\theta(x,x')$.
		\State $\delta\gets V_\psi(x)+r(x,x')-\bar{r}_n-V_\psi(x')$
		\State $\theta\gets\theta+\alpha^\theta_n\delta\nabla_\theta \ln P_\theta(x|x')$
		\State $\psi\gets\psi+\alpha^\psi_n\delta\nabla_\psi V_\psi(x')$
		\State $\bar{r}\gets\bar{r}+\alpha^R_n\delta$
		\State $n\gets n+1$
		\Until{$n=N$}
	\end{algorithmic}
\end{algorithm}

This algorithm, and the one discussed in the next section, can be extended in many of the ways previously discussed in section \ref{sec:review}.
Further, it can be manipulated to an approximate form which more closely matches the non actor-critic algorithms of section \ref{dynamical_gradients}.
To see this, we consider modifying the algorithm to use an $n$-step update with extremely large $n$: in this case, the value function can be removed, as its contribution from the target $n$ step state averages out over the stationary state to zero when $n$ is sufficiently large.
The resulting algorithm is equivalent to a continuing version of algorithm \ref{monte_carlo_reinforce_value_baseline}.
Further, the current state value is simply a baseline which can be removed, producing an algorithm  equivalent to that used in \cite{Das2019}, a continuing version of algorithm \ref{monte_carlo_reinforce}.
This provides an approximate, value-free algorithm for the continuing case.
Alternatively, the algorithm in \cite{Das2019} can be seen as making a finite time approximation to the problem itself, using algorithm \ref{monte_carlo_reinforce} of the previous section with an additional average reward baseline.

\subsection{An approximate approach: discounting}
The more traditional approach in RL for continuing problems gets round the issue of divergent returns by discounting the contribution of rewards to the value of a state proportional to how long after the state the reward was given.
That is, the value of a state is defined as
\begin{eqnarray}\label{discounted_value}
V^\gamma_{P_\theta}(x)=\lim_{T\rightarrow\infty}\left\langle\sum_{t'=t}^T\gamma^{t'-t}r(x_{t+1},x_t)\right\rangle_{P_\theta,X_t=x},
\end{eqnarray}
which is convergent for a discount rate $\gamma$ less than $1$.

For these values to be correct, the discounting must be introduced in the original definition of the problem: in this case, the interpretation of the discount is a probability of the system entering an absorbing state in which it receives no more reward \cite{Thomas2014}.
Sampling states correctly then takes us back to a finite trajectory based approach, where we initialize according to some distribution, and end the trajectory at some variable time with probability $1-\gamma$ at each time step, causing infinite trajectories to be exponentially suppressed.

While this may be an interesting problem in its own right, this is not the problem we are aiming to solve.
Instead, we introduce discounted values as an approximate approach to estimating the dynamical gradient for the average return problem outlined in the previous section.
This allows us to cease tracking the average return, while often providing lower variance estimates for the gradient, at the expense of accuracy in the final result.

For this approximate approach to produce reasonable accuracy of the final result, theoretical work in the RL literature has suggested that the discount rate $\gamma$ must be such that $1/(1-\gamma)$ -- the time-scale for the average time between transitions to the absorbing state -- is larger than the mixing time of the current dynamics $P_\theta$ \cite{Kakade2001,Bartlett2002,Marbach2003,Thomas2014}.

To gain an intuition for why discounting works for large enough values, lets consider a slightly modified definition of the differential values.
Truncating our earlier definition up to a finite time, we use the return up to that time averaged over time and an initial stationary distribution
\begin{eqnarray}
\bar{r}^T_\theta=\frac{1}{T}\left\langle R(\omega_{0}^T)\right\rangle_{P_\theta}=\sum_{x_0}P^\mathrm{ss}_\theta(x_0)\left\langle R(\omega_{0}^T)\right\rangle_{P_\theta,X_0=x_0},
\end{eqnarray}
where $\lim_{T\rightarrow\infty}\bar{r}_\theta^T=\bar{r}_\theta$.
We negate this average off the reward at each step to define our truncated differential values, finding
\begin{eqnarray}
V^T_{P_\theta}(x_0)
&=\left\langle R(\omega_{0}^T)\right\rangle_{P_\theta,X_0=x_0}-T\bar{r}_\theta^T\nonumber\\
&=\sum_{t=1}^T\sum_{x_t,x_{t-1}}P_\theta(x_t,x_{t-1})r(x_t,x_{t-1})\left[P_\theta(x_{t-1}|x_0)-P^\mathrm{ss}_\theta(x_{t-1})\right],
\end{eqnarray}
where in the second line we have split the returns in to reach reward, summing over the possible paths up to each pair, with $P_\theta(x_{t-1}|x_0)$ used to represent the probability of reaching $x_{t-1}$ under $P_\theta$ by any path initiated from $x_0$. 
Introducing an importance sampling factor, we may then rewrite the value function in terms of a return in which the rewards depend on the state being valued: given
\begin{eqnarray}
R'(\omega_{0}^T)=\sum_{t=1}^Tr(x_t,x_{t-1})\frac{P_\theta(x_{t-1}|x_0)-P^\mathrm{ss}_\theta(x_{t-1})}{P_\theta(x_{t-1}|x_0)},
\end{eqnarray}
we have
\begin{eqnarray}
 V^T_{P_\theta}(x_0)
 &=\left\langle R'(\omega_{0}^T)\right\rangle_{P_\theta,X_0=x_0}.
\end{eqnarray}

While this equation requires no knowledge of the average return, it does require extremely detailed knowledge of the probabilities of states conditioned on states multiple steps in the past, something not easily accessible.
However, this form makes it transparent that by negating the average return, we are essentially decaying out the contribution of rewards received many steps in the future, in a fashion reminiscent of discounting: since we assume ergodicity, as the time after valuation extends into the future the conditional probability will converge to the stationary sate.

To see this decay we use a spectral decomposition of an operator which describes the evolution of probability distributions under the dynamics $P_\theta$.
Viewing $P_\theta(x|x')$ as the components of a transition matrix describing the evolution of a probability distribution
\begin{eqnarray}
\mathcal{W}_\theta=\sum_{x,x'}P_\theta(x|x')\ket{x}\bra{x'}.
\end{eqnarray}
This matrix can be diagonalized, resulting in left $\bra{l_i}$ and right $\ket{r_i}$ eigenvectors
\begin{eqnarray}
\bra{l_i}&=\sum_x l_i(x)\bra{x}, \qquad \ket{r_i}&=\sum_x r_i(x)\ket{x},
\end{eqnarray}
which are orthogonal, $\left\langle l_i | r_j \right\rangle=\delta_{ij}$, with eigenvalues $\lambda_i$ satisfying $\bra{l_i}\mathcal{W}_\theta=\lambda_i\bra{l_i}$ and $\mathcal{W}_\theta\ket{r_i}=\lambda_i\ket{r_i}$.
The stationary state satisfies $\mathcal{W}_\theta\ket{P_\theta^\mathrm{ss}}=\ket{P_\theta^\mathrm{ss}}$, corresponding to an eigenvalue of $1$, with associated left eigenvector the ``flat'' state $\bra{-}$ with value $1$ for every component.
It can further be shown that all eigenvalues will satisfy $|\lambda_i|<1$, since we are assuming the model is ergodic and thus has a single stationary state.

Given this spectrum, we may expand the time evolution of a given initial probability distribution as
\begin{equation}
\ket{P(t)}=\mathcal{W}_\theta^t\ket{P(0)}=\ket{P_\theta^\mathrm{ss}}+\sum_{i=2}^D\lambda^t_i\ket{r_i}\left\langle l_i | P(0)\right\rangle,
\end{equation}
where $D$ is the dimension of the state space.
This allows us to rewrite the probabilities $P_\theta(x_{t-1}|x_0)$ in a spectral expansion, by taking as our initial distribution $\ket{P(0)}=\ket{x_0}$ and projecting out the $x_{t-1}$ component
\begin{eqnarray}
P_\theta(x_{t-1}|x_0)=P_\theta^\mathrm{ss}(x_{t-1}) + \sum_{i=2}^D \lambda_i^{t-1} r_i(x_{t-1}) l_i(x_0).
\end{eqnarray}

Finally, substituting this into our alternative equation for the truncated values, we have
\begin{eqnarray}
R'(\omega_{1}^T,x_0)=\sum_{t=1}^Tr(x_t,x_{t-1})\frac{\sum_{i=2}^D \lambda_i^{t-1} r_i(x_{t-1}) l_i(x_0)}{P_\theta(x_{t-1}|x_0)}.
\end{eqnarray}
Recalling $|\lambda_i|<1$ for $i\neq1$, all terms in this sum decay as time increases, and thus later rewards contribute less and less to the differential return.
For later times this decaying contribution is dominated by the leading eigenvalue of the master operator, the inverse of the relaxation time of the Markov chain, with the denominator becoming the stationary distribution
\begin{eqnarray}
R'(\omega_{1}^T,x_0)\approx\sum_{t=1}^Tr(x_t,x_{t-1})\frac{\lambda_2^{t-1} r_2(x_{t-1}) l_2(x_0)}{P_\theta^\mathrm{ss}(x_{t-1})}.
\end{eqnarray}
This is suggestive of the form of return used when discounting, with some similarity between $\lambda_2$ and the discount $\gamma$: indeed, the mixing time, which $1/(1-\gamma)$ must be less than for accuracy, is closely related to the relaxation time of the dynamics given by $1/(1-\lambda_2)$.

Replacing all of the above probabilities with a general discounting factor is clearly an approximation of the true differential values, and thus introduces a bias in the final results.
However, it removes the need to track the average return in order to estimate the temporal differences, which can itself introduce errors and bias into the optimization.
Discounting can also lower variance of the gradient estimate, as discounting reduces the impact of stochasticity by giving less weight to the further future.
As such, we now detail how to use discounted values to guide the evolution of the dynamical weights.

To optimize an approximation for the discounted values, we note that the values of equation \eref{discounted_value} satisfy a slightly modified Bellman equation
\begin{eqnarray}\label{discounted_bellman}
V_{P_\theta}(x')=\sum_{x}P_\theta(x|x')\left[\gamma V_{P_\theta}(x)+r(x,x')\right].
\end{eqnarray}
We thus follow the same semi-gradient approach as previously, using the gradient estimate
\begin{eqnarray}\label{discounted_value_loss}
\fl\nabla_\psi L_V(\psi)\approx-\sum_{x,x'}P_\theta(x|x')P_\theta^\mathrm{ss}(x')\left[\gamma V_\psi(x)+r(x,x')-V_\psi(x')\right]\nabla_\psi V_\psi(x'),
\end{eqnarray}
given by the discounted temporal difference error
\begin{eqnarray}
\delta_{\gamma\mathrm{TD}}(x,x')=\gamma V_{\psi}(x)+r(x,x')-V_{\psi}(x').
\end{eqnarray}
To approximate the dynamical gradient, we use this temporal difference as an approximation to the one appearing in equation \eref{differential_dynamics_loss}, arriving at
\begin{eqnarray}\label{discounted_dynamics_loss}
\fl\nabla_\theta d_{KL}(P_\theta|P_W)
\approx-\sum_{x,x'}P_\theta(x|x')P_\theta^\mathrm{ss}(x')\left[\gamma V_{\psi}(x)+r(x,x')-V_{\psi}(x')\right]\nabla_\theta \ln P_\theta(x|x').\nonumber\\
\end{eqnarray}
The resulting online algorithm \ref{discounted_actor_critic}, almost identical to the one for differential returns, is given below.

\begin{algorithm}[h]
	\caption{KL regularized discounted actor-critic}\label{discounted_actor_critic}
	\begin{algorithmic}[1]
		\State \textbf{inputs} dynamical approximation $P_\theta(x,x')$, value approximation $V_\psi(x)$
		\State \textbf{parameters} learning rates $\alpha^\theta_n$, $\alpha^\psi_n$; total updates $N$, discount factor $\gamma$
		\State \textbf{initialize} choose initial weights $\theta$ and $\psi$, define iteration variable $n$, individual error $\delta$
		\State $n\gets0$
		\Repeat
		\State Generate a transition from $x'$ to $x=\{x,F(x,x')\}$ according to the dynamics given by $P_\theta(x,x')$.
		\State $\delta\gets \gamma V_\psi(x)+r(x,x')-V_\psi(x')$
		\State $\theta\gets\theta+\alpha^\theta_n\delta\nabla_\theta \ln P_\theta(x|x')$
		\State $\psi\gets\psi+\alpha^\psi_n\delta\nabla_\psi V_\psi(x')$
		\State $n\gets n+1$
		\Until{$n=N$}
	\end{algorithmic}
\end{algorithm}

\subsection{Infinite horizon example: random walker on a ring}
As a simple example to demonstrate both these algorithms, we return to our particle hopping on a chain example, making the chain periodic with length $L$, $x\in{0,...,L-1}$.
The original dynamics we consider is inspired by a model in Ref.~\cite{Ferre2018,Das2019}.
We take a dynamics given by a periodic potential, specifically
\begin{eqnarray}
	P(x+1|x)=\sigma\left(u+v\sin\left(\frac{2\pi x}{L}\right)\right),
\end{eqnarray}
where $\sigma(y)=e^y/(1+e^y)$ is the sigmoid function, and $u,v$ are parameters of the dynamics.
Our goal is to study rare trajectories of the particles transition direction, with the sign of the bias $s$ determining whether we focus on trajectories where the direction moved is largely positive or negative.
To achieve this we introduce a soft condition by weighting transitions as
\begin{equation}
W(x,x')
=\left\{\begin{array}{l@{\qquad}l}
e^{-s}  & (x'-1)\:\mathrm{mod}\: L=x\\
e^{s}  & \mathrm{otherwise}
\end{array}\right..
\end{equation}

For function approximations, we could choose a tabular approach as we did for the excursions, which would work perfectly well in this simple scenario.
To demonstrate a more sophisticated function approximation, making the algorithms learn faster while requiring less data, here we instead choose to use a linear expansion in set of Fourier modes.
That is, we set the dynamics to $P_\theta(x+1|x)=\sigma\left(U(x)\right)$ with potential
\begin{eqnarray}\label{linear_probabilities}
	U(x)=\sum_i\theta_i f_i(x),
\end{eqnarray}
where each $f_i$ is chosen to be either a fourier mode or the flat function $f_i(x)=1$, and the values are set to
\begin{eqnarray}
	V_\psi(x)=\sum_i\psi_i f_i(x),
\end{eqnarray}
for the same set of functions $f_i$.
The gradients of these approximations are closely related to the values of this ``feature vector'' $\vec{f}$, with
\begin{eqnarray}
	\nabla_\psi V_\psi(x) = \vec{f}(x),
\end{eqnarray}
and
\begin{eqnarray}
	\nabla_\theta \ln P_\theta\left(x\pm1|x\right)=\pm\vec{f}(x)P_\theta\left(x\mp1|x\right).
\end{eqnarray}
We train these approximations using both the differential and discounted forms of AC, annealing the bias $s$ across a range of values.
By initiating the weights from those found training at nearby values of the bias, we can potentially reduce the number of updates required to achieve good results.

\begin{figure}
	\begin{center}
		\includegraphics[width=1\linewidth]{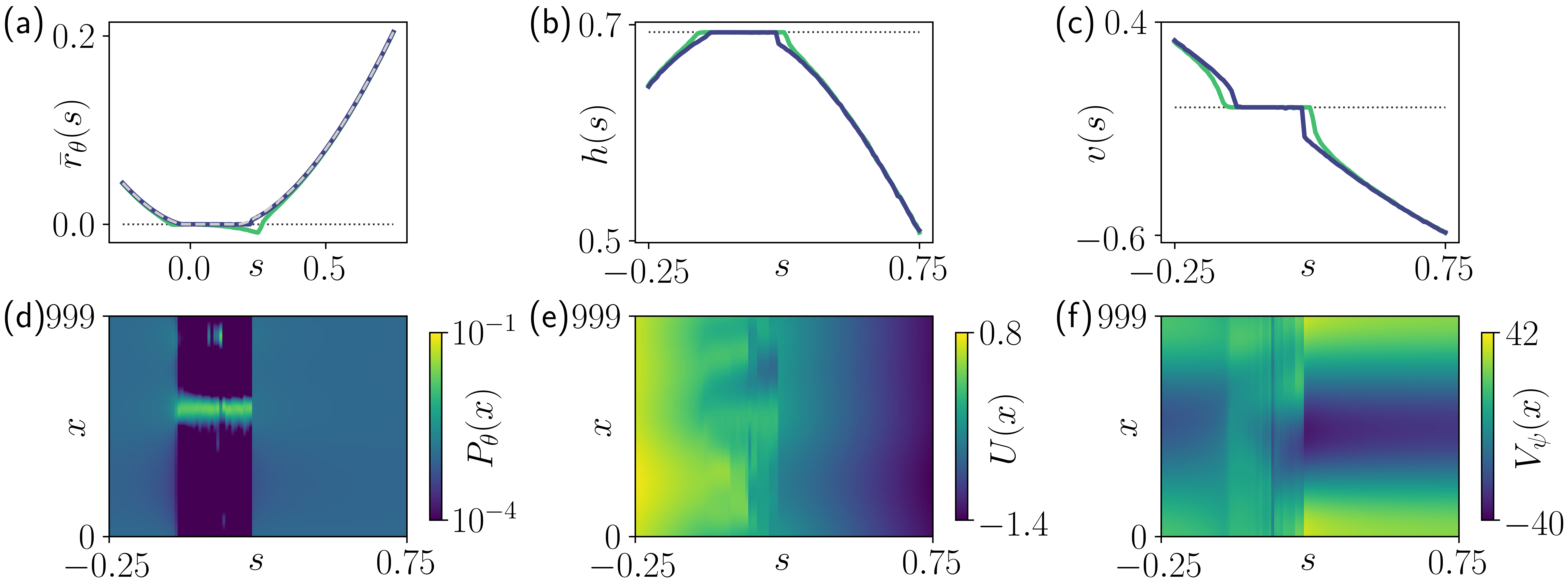}
	\end{center}
	\caption{\label{fourier_ring} \textbf{Fourier-expansion ring}. Results for a ring of length $L=1000$, with $u=0.15$ and $v=0.3$. In plots (a-c), the dark (purple) lines are results produced using differential actor-critic, while light(green) lines are for discounted actor-critic with a discount of $\gamma=0.999$. Dotted gray lines show the values at $s=0$. Plots (d-f) Show results for the differential actor-critic. (a) Time average of the rewards received each transition, i.e.\ the scaled cumulant generating function for this observable, as a function of the bias. The dashed gray line indicates the exact result calculated according to \hyperref[doob-dynamics]{Appendix C}. (b) The entropy of the dynamics. (c) The time-average of the current, the observable biased against. (d) The steady-state distribution of the learnt dynamics as a function of $s$. (e) The potential $U(x)$ defining the probability of going up, $P_\theta(x+1|x)$, learnt for each $s$. (f) The value of each state found during training.}
\end{figure}

Results are shown in figure \ref{fourier_ring}, with the first row showing: (a) the time-averaged reward $\bar{r}_\theta$; (b) an estimate of the entropy of the dynamics, defined by
\begin{eqnarray}
	h=-\sum_{x,x'}P_\theta(x'|x)P_\theta(x)\ln P_\theta(x'|x);
\end{eqnarray}
(c) an estimate of the time-averaged current
\begin{eqnarray}
	v=\frac{1}{s}\sum_{x,x'}P_\theta(x'|x)P_\theta(x)\ln W(x,x'),
\end{eqnarray}
with exact results for the time-averaged reward calculated for comparison as described in \hyperref[doob-dynamics]{Appendix C}.
As can be seen from plot \ref{fourier_ring}(a), the differential AC provides results with a high degree of accuracy, while the discounting appears to be inaccurate near transitions in the trajectory statistics.
This is as expected: near a transition, long-time correlations will become important to the statistics, and discounting puts a cap on how much of the future is taken into account.
Figure \ref{fourier_ring}(d) shows the steady state-distribution across the ring, with a region of localization occurring for values of positive bias which are not enough for the optimized dynamics to overcome the constant force of the model.
Despite the low entropy of the steady state caused by this localization, this range of biases is in fact where the entropy of the dynamics is highest: here, transitions are likely to occur either up or down, causing the localization.
Outside this range the majority of transitions are either up or down, depending on the sign of the bias.
The potential defining the probability of going up, the term inside the sigmoid of equation \eref{linear_probabilities}, is shown in figure \ref{fourier_ring}(e), with 0 causing equal probability of up or down.
Outside the range of biases resulting in localization, we find a clear favour towards going in a direction prescribed by the bias, with the potential either taking significant positive or negative values.
Inside the localized range, the potential has an oscillatory structure, which we note will only be accurate where the stationary state is non-negligible.

\subsection{Connection to large deviation cumulant generating functions}\label{sec:ld-scgf}
The construction used in this section is closely related to the theory of large deviations, as should be expected given recent connections between the large deviations of trajectories and optimal control theory \cite{Jack2015a,Chetrite2015}.
The optimal dynamics for minimizing the time averaged KL divergence is in fact the the dynamics resulting from the generalised Doob transformation \cite{Jack2010,Chetrite2015}.
Additionally, the long-time average of the logged partition function $z$ of equation \eref{scgf} is exactly the scaled cumulant generating function (SCGF), the Legendre transform of which provides the probability distribution of the observable whose rare events we are studying.
Rearranging equation \eref{average_return}, we have
\begin{eqnarray}
z=\bar{r}_\theta+d_{KL}(P_\theta|P_W),
\end{eqnarray}
which holds for any dynamics $P_\theta$.
While the KL divergence part of this equation is difficult to calculate, our algorithms are designed to minimize this term, approaching zero at optimality. 
While optimizing we can easily calculate $\bar{r}_\theta$: indeed, this is already a part of the differential AC algorithm.
Thus, these algorithms provide direct access to the SCGF, and therefore the statistics of the rare events.

Minimizing the KL divergence is equivalent to maximizing the return, and since the KL divergence is non-negative we may rewrite
\begin{eqnarray}
z&\geq\bar{r}_\theta\quad\forall\quad\theta\nonumber\\
&\geq\max_\theta\bar{r}_\theta\nonumber\\
&\geq\max_\theta \sum_{x',x}P_\theta^\mathrm{ss}(x)P_\theta(x'|x)\left[\ln W(x',x) -
\ln\left(\frac{P_\theta(x'|x)}{P(x'|x)}\right)\right],
\end{eqnarray}
with the inequality saturable if the Doob dynamics is contained within the variational space spanned by $\theta$ for the chosen function approximation, that is
\begin{eqnarray}
	z=\max_{\tilde{P}}\sum_{x',x}\tilde{P}^\mathrm{ss}(x)\tilde{P}(x'|x)\left[\ln W(x',x) -
	\ln\left(\frac{\tilde{P}(x'|x)}{P(x'|x)}\right)\right],
\end{eqnarray}
as seen in the LD literature discussing connections to optimal control \cite{Jack2015a,Chetrite2015}.
The time-averaged reward estimated during training thus provides an efficient way of calculating at least a lower bound of the SCGF, with powerful function approximations and extensive training allowing access to an accurate value without needing to use any other form of statistical sampling.
In cases where high degrees of accuracy are not possible, the learnt dynamics can be combined with sampling techniques such as TPS or cloning to calculate a better estimate.

\section{Conclusions and outlook}\label{generalizations}
\label{conclusion}
In this work we have highlighted a general approach for developing numerical approaches to study questions about statistical ensembles of trajectories, with a particular focus on ensembles consisting of rare trajectories of some original dynamics.
We have shown that gradient based optimization of a sampling dynamics for these trajectory ensembles naturally maps onto a regularized form of reinforcement learning, closely related to maximum-entropy reinforcement learning.
We used this connection to pedagogically develop algorithms in a finite time setting, a key ingredient being the extensive use of value functions, a first in the rare trajectory sampling literature.
Reviewing a range of modifications to learning algorithms and choices of function approximations found in the reinforcement learning literature, we saw just how many possibilities this connection makes available for the study of rare trajectories.
We then adapted the approach for time-homogeneous problems which have no unique time and can be viewed as single unending trajectories, for the study of statistics of time-averaged observables, and described how this connects to the theory of large deviations for Markov processes and its relationship with optimal control theory.
This development was supplemented by two examples: generating random walker excursions with the correct probabilities for the finite time case, and statistics of the time-averaged current for a particle on a ring in the infinite time case.

There is a wide range of possible avenues for future research building on what we have presented here. 
An obvious one is using these algorithms to tackle more sophisticated problems than the simple models we used as illustration.
For example, we may seek to apply the approach to study rare trajectories of many-body systems such as spin lattices or molecular dynamics, where the state space grows exponentially with the number of particles.
In this situation, the algorithms are essentially unchanged: the difficulty comes in making an appropriate choice of function approximation, such that it can efficiently encode the dynamics.
Analytical study of many problems can produce simple, physically inspired parameterizations of the dynamics in such many body systems, see e.g. \cite{Oakes2018,Whitelam2019,Dolezal2019}.
Where these physically inspired approximations cease to be sufficient, or where it is difficult to gain such insight, we could instead resort to neural networks.
These have proven to be an incredibly versatile function approximations, with extreme representative power.
Their application to RL comes with a caveat, however: they are unstable with the simpler algorithms we have presented.
As discussed in section \ref{neural-networks}, to overcome these issues, training of neural networks must therefore be conducted using modified algorithms.
Further examples of the use of neural networks in large deviations can be found in \cite{Oakes2020,Casert2020}.

Many-body problems will bring with them a separate issue to overcome: how to achieve sufficiently broad sampling of the state space, especially in models near phase transitions, where Markov chain sampling can become trapped in subsets of the state space.
The trapping could lead to over fitting of the function approximation on the current area of the state space the Markov chain is sampling, forgetting the dynamics in previously visited regions.
This is a problem which may be addressed by running multiple trajectories in parallel, or through the use of replay buffers to further sample previously visited regions of the state space.

Beyond applications, interesting generalizations and extensions include:
\begin{itemize}
	\item \textbf{Limited control.} In certain situations it may be beneficial (or only possible) to make part of the dynamics adaptive. For example, in a many-body system where each particle has separate degrees of freedom such as a position and orientation, we may only control the orientational evolution while leaving the position unchanged from the original dynamics. In this setup, the evolution of the position takes on the role of an environment from the RL perspective, with the orientation under the control of the agent. While this may limit the effectiveness of the resulting dynamics for sampling, it could be much easier to optimize, requiring less parameters or having a more obvious choice of function approximation.
	
	\item \textbf{Non-Markovian original dynamics.} As discussed earlier, the approach developed in this work can be almost immediately extended to arbitrary non-Markovian original dynamics in the finite time case. For example, the Monte Carlo returns with a value baseline becomes based on the gradients
	\begin{eqnarray}\label{eq:non-markovian-dynamics}
	\fl \nabla_\theta D_{KL}(P_\theta|P_W)
	=-\left\langle\sum_{t=1}^{T}\left(R_W(\omega_0^T)-V_\psi(\omega_0^t)\right)\nabla_\theta\ln P_\theta(x_{t}|\omega_{0}^{t-1})\right\rangle_{P_\theta},
	\\
	\fl \nabla_\psi L_V(\psi)
	=-\left\langle\sum_{t=1}^{T}\left(R_W(\omega_0^T)-V_\psi(\omega_0^t)\right)\nabla_\psi V_\psi(\omega_0^t)\right\rangle_{P_\theta},
	\end{eqnarray}
	where we have simply replaced the state and time with the full history of the trajectory, sampling with a parameterized dynamics which is itself non-Markovian. 
	While general, this is more likely to be applicable with approximation in studying the statistics of problems where the original dynamics has a limited amount of memory. 
	An alternative use case is a side effect of using function approximations: since some useful information may be lost in processing the state, the dynamics is effectively non-Markovian.
	Making use of processed states, i.e. feature vectors, of a recent history of states may thus improve the accuracy of the dynamics further.
	A similar modification can be made for the infinite time case when the original dynamics has a limited range of non-Markovianity, or the weighting depends on a short part of the history of previous states.
	A particularly powerful function approximation to apply in such problems is that of recurrent neural networks.
	
	\item \textbf{Non-Markovian weights.} Rather than the original dynamics being non-Markovian, its possible that the weights may be non-Markovian.
	That is, rather than taking the transition-local product structure of \eref{eq:product-weighting}, the weight of each trajectory may simply be some function $W\left(\omega_0^T\right)$.
	Generically this will result in a problem identical in structure to the one above in \eref{eq:non-markovian-dynamics}: even if the original dynamics remains Markovian, the non-Markovian nature of the weights will necessitate a non-Markovian parameterized dynamics to best sample the reweighted ensemble.
	However, many non-Markovian weights may only require a subset of the information contained in the trajectories history.

	For example, suppose we wish to consider the subset of random walks with a particular area $A$.
	To do this we would set the weights to
	\begin{eqnarray}
		W\left(\omega_0^T\right)=\delta_{A,A\left(\omega_0^T\right)}.
	\end{eqnarray}
	There is no obvious way to split this weight up, but we can observe that the only information about the history necessary to calculate the weight at the end is its area $A\left(\omega_0^T\right)$.
	As such, as a trajectory evolves the only information we need keep track of is the area up to each time $A_t$, updating it after each transition.
	It seems reasonable that the optimal dynamics to sample this ensemble may only be conditional on only the current state, time, and the area up to that point in the trajectory: that is, it should be sufficient to parameterize a conditional dynamics $P_\theta(x_{t+1}|x_{t},t,A_t)$.
	This can in fact be proven, and presents a particular case of what we call a \textbf{generalized state}: the necessary information, in this case $(x_t,t,A_t)$ from the trajectories history to be able to exactly reproduce the reweighted ensemble.
	In future work, we will further expand on the idea of generalized states, applying our approach to more complex conditional problems.

	\item \textbf{Fluctuating time ensembles.} Rather than ending trajectories at a fixed time, we could end trajectories according to some condition, for example, to study the statistics of rare first passages. Given that variable length trajectories are the natural setting of reinforcement learning, these algorithms will have natural adaptations to sampling in these problems, with optimal sampling dynamics likely being time-independent.
	
	\item \textbf{Continuous time Markov processes:} Here for concreteness we presented our approach 
	for discrete-time dynamics, but it can easily be generalised to both continuous-time jump processes, to diffusions, and to combinations of both. 
	Indeed, there is already an extensive literature of work covering continuous time versions of reinforcement learning \cite{Bradtke1994,Doya2000,Munos2005,Vamvoudakis2010,Fremaux2013}.
	In fact, the continuous time version of our loss-functions have already been discussed in \cite{Chetrite2015}, where connections were made between large deviation theory and control theory.
	Further to this, there has already been some adaptive algorithms of a similar nature developed for sampling rare trajectories in the continuous time case. In particular, \cite{Das2019} uses an algorithm which is an approximation to an ``$\infty$-step'' version of the differential actor-critic algorithm 
	described above. 
	This allows the removal of the value function, since for the current state it is a baseline, and for the potential ``$\infty$-step'' states the value averages to zero over the stationary state.
	Approximations result from truncating the partial return between these two times to a finite length. Additionally, in \cite{Kappen2016} the KL divergence is used with the parameterized and weighted distributions swapped around.
	Finally, a version of the LSTD algorithm \cite{Bradtke1996,Sutton2018} applied to the non-linear Bellman equation \eref{1-step_gauge} \cite{Hasselt2019} has recently been developed for large deviations of diffusive systems \cite{Ferre2018}.
	Despite the above developments, value functions and the many other techniques found in RL are not currently used for the sampling of rare, continuous time trajectories.
	
	\item \textbf{Use in TPS or cloning:} If the function approximation is incapable of achieving a sufficient accuracy to study the rare events (e.g. to directly estimate the SCGF using optimized trajectories) then TPS or cloning could be used to fix the statistics, with convergence sped up by the optimized dynamics \cite{Oakes2018,Oakes2020, Das2019}.
\end{itemize}

Beyond these applications of RL-like techniques to statistical sampling, there is the obvious potential of taking this connection in the other direction, to gain further understanding of RL itself through the use of techniques and intuitions from the statistical physics perspective.

\section*{Acknowledgements}
The authors thank A. Lamacraft for bringing to our attention literature central to the development of this project.
We also thank the referees for comments leading to a significant improvement in presentation.
This research was funded in part by University of Nottingham grant no.\ FiF1/3 and The Leverhulme Trust grant no.\ RPG-2018-181. We are grateful for access to the University of Nottingham Augusta HPC service. We also acknowledge the use of Athena at HPC Midlands+, which was funded by the EPSRC on grant EP/P020232/1 as part of the HPC Midlands+ consortium.

\appendix

{\renewcommand{\addtocontents}[2]{} \section{Exact optimal sampling and random walk excursions}}
\addcontentsline{toc}{section}{Appendix A~~Exact optimal sampling and random walk excursions}
\label{exact-excursions}
In this appendix we demonstrate how the optimal dynamics can be calculated exactly, either analytically or numerically.
This is done by propagating an iterative equation for a function of the state and time, which is used to rescale the original transition probabilities.
While in principle this can solve any problem, it can be numerically unstable, and will not be applicable as presented to problems which are the target application of the current line of research: systems for which the state space is too large for a single value to be associated to every state.
It is expected that these techniques can also be extended to generic function approximation (see Ref. \cite{Ferre2018} for linear approximations in diffusion processes), however, it is likely less stable than algorithms based on the KL divergence, due to multiplicative (rather than additive) nature of the objects involved frequently causing extremely large or small numerical values.

Beginning from
\begin{eqnarray}
P_W\left(\omega_{0}^T\right)=\frac{\prod_{t=0}^TW(x_t,x_{t-1},t)\prod_{t=1}^TP(x_t|x_{t-1})P(x_0)}{\sum_{\omega_0^T}W\left(\omega_0^T\right)P\left(\omega_{0}^T\right)},
\end{eqnarray}
we aim for a time dependent Markovian dynamics generating this ensemble.
However, rather than assuming this is possible, we first calculate a decomposition into non-Markovian conditional probabilities, producing
\begin{eqnarray}
	P_W\left(\omega_{0}^T\right)=\prod_{t=0}^TP_W(x_t|\omega_0^{t-1}).
\end{eqnarray}
To do this, we use the definition of a conditional probability in terms of joint probability distributions: iterating backwards step by step we have
\begin{eqnarray}
	P_W\left(\omega_{0}^{t-1}\right)=\sum_{x_t}P_W\left(\omega_{0}^{t}\right),
\end{eqnarray}
and thus
\begin{eqnarray}
	P_W\left(x_t|\omega_{0}^{t-1}\right)=\frac{P_W\left(\omega_{0}^{t}\right)}{P_W\left(\omega_{0}^{t-1}\right)}.
\end{eqnarray}
Combining these definitions, for the final timestep we have
\begin{eqnarray}
P_W\left(x_T|\omega_{0}^{T-1}\right)
&=\frac{\prod_{t=0}^TW(x_t,x_{t-1},t)P\left(\omega_{0}^T\right)}{\sum_{x_T}\prod_{t=0}^TW(x_t,x_{t-1},t)P\left(\omega_{0}^T\right)}\nonumber\\
&=\frac{W(x_T,x_{T-1},T)P\left(x_T|x_{T-1}\right)}{\sum_{x_T}W(x_T,x_{T-1},T)P\left(x_T|x_{T-1}\right)}\nonumber\\
&=\frac{W(x_T,x_{T-1},T)P\left(x_T|x_{T-1}\right)}{\mathbb{E}_{x_T\sim P}\left[W(x_T,x_{T-1},T)|x_{T-1}\right]},
\end{eqnarray}
where we see that despite starting from joint probabilities over the whole history of the trajectory, the end result is invariant over all but the state prior to the transition, and thus we may write $P_W\left(x_T|\omega_{0}^{T-1}\right)=P_W\left(x_T|x_{T-1},T\right)$ for all past trajectories up to the final transition.
For earlier times we have
\begin{eqnarray}
\fl P_W\left(x_t|\omega_{0}^{t-1}\right)
&=\frac{\sum_{\omega_{t+1}^T}\prod_{t'=0}^TW(x_{t'},x_{t'-1},t')P\left(\omega_{0}^T\right)}{\sum_{\omega_{t}^T}\prod_{t'=0}^TW(x_{t'},x_{t'-1},t')P\left(\omega_{0}^T\right)}\nonumber\\
&=\frac{\left[\sum_{\omega_{t+1}^T}\prod_{t'=t+1}^TW(x_{t'},x_{t'-1},t')P\left(\omega_{t+1}^T|x_{t}\right)\right]W(x_{t},x_{t-1},t)P\left(x_t|x_{t-1}\right)}{\sum_{\omega_{t}^T}\prod_{t'=t}^TW(x_{t'},x_{t'-1},t')P\left(\omega_t^T|x_{t-1}\right)}\nonumber\\
&=\frac{\mathbb{E}_{\omega_{t+1}^{T}\sim P}\left[\prod_{t'=t+1}^TW(x_{t'},x_{t'-1},t')|x_t\right]W(x_{t},x_{t-1},t)P\left(x_t|x_{t-1}\right)}{\mathbb{E}_{\omega_{t}^{T}\sim P}\left[\prod_{t'=t}^TW(x_{t'},x_{t'-1},t')|x_{t-1}\right]},
\end{eqnarray}
where similarly to the final transition, the dependence on the past prior to the state before the transitions at each time have cancelled out, allowing us to write $P_W\left(x_t|\omega_{0}^{t-1}\right)=P_W\left(x_t|x_{t-1},t\right)$ for all times.
Finally, the initial distribution is modified as
\begin{eqnarray}
P_W(x_0)
&=\frac{\mathbb{E}_{\omega_{1}^{T}\sim P}\left[\prod_{t'=1}^TW(x_{t'},x_{t'-1},t')|x_t\right]W(x_{0},0)P\left(x_0\right)}{\mathbb{E}_{\omega_{0}^{T}\sim P}\left[\prod_{t'=0}^TW(x_{t'},x_{t'-1},t')\right]}.
\end{eqnarray}

These expectations represent the expected contribution to the weighting of the trajectories future given the current state and time.
The individual contributions to the expectation play a similar role to the returns in our algorithms, however, now they have a product structure over the individual factors associated to each transition, rather than a sum structure.
Labeling these expectations as
\begin{eqnarray}\label{distributed_expectation}
g(x_t,t)=\mathbb{E}_{\omega_{t+1}^{T}\sim P}\left[\left.\prod_{t'=t+1}^TW(x_{t'},x_{t'-1},t')\right|x_t\right],
\end{eqnarray}
with $g(x,T)=1$ for all $x$, we have
\begin{eqnarray}\label{distributed_optimal}
P_W\left(x_t|x_{t-1},t-1\right)=\frac{g(x_t,t)}{g(x_{t-1},t-1)}W(x_{t},x_{t-1},t)P\left(x_t|x_{t-1}\right),
\end{eqnarray}
for all $t$. 
The function $g$, related to a gauge transformation of the trajectory probabilities, can then be efficiently calculated by iterating backwards, using
\begin{eqnarray}\label{distributed_gauge}
g(x_t,t)=\mathbb{E}_{x_{t+1}\sim P}\left[W(x_{t+1},x_{t},t+1)g(x_{t+1},t+1)\right].
\end{eqnarray}

\textbf{Excursions.}
We now demonstrate the above approach by calculating the transformation for the conditioned random walk excursions case mentioned in section \ref{background}.
This problem possesses a lightcone structure inherited from the original random walker dynamics: since each transition can only go up or down one, the position $n$ steps in the future or past can only be $n$ higher or lower than the present position.
Since we are targeting a dynamics which will entirely end in a single state, this lightcone structure means the backwards iteration based on Eq. \eref{distributed_gauge} will simplify significantly, allowing analytical solution.

With the weights defined by $W(x',x,T)=\delta(x')$ and $W(x',x,t)=H(x')$ we have
\begin{eqnarray}
g(x,t)=\frac{1}{2}\left(H(x+1)g(x+1,t+1)+H(x-1)g(x-1,t+1)\right),
\end{eqnarray}
for $t<T-1$, with end condition $g(x,T)=1$ for all $x$ and
\begin{eqnarray}
g(x,T-1)=\frac{1}{2}\left(\delta(x+1)+\delta(x-1)\right).
\end{eqnarray}
This immediately implies that $g(x,t)=0$ if $x<-1$ from the heaviside step function, and $g(-1,t)=0.5g(0,t)$ on the positive-negative boundary.
The lightcone structure, imposed by the delta function at the final time, results in $g(x,t)=0$ for $x>T-t$.

\begin{figure}
	\begin{center}
		\includegraphics[width=0.7\linewidth]{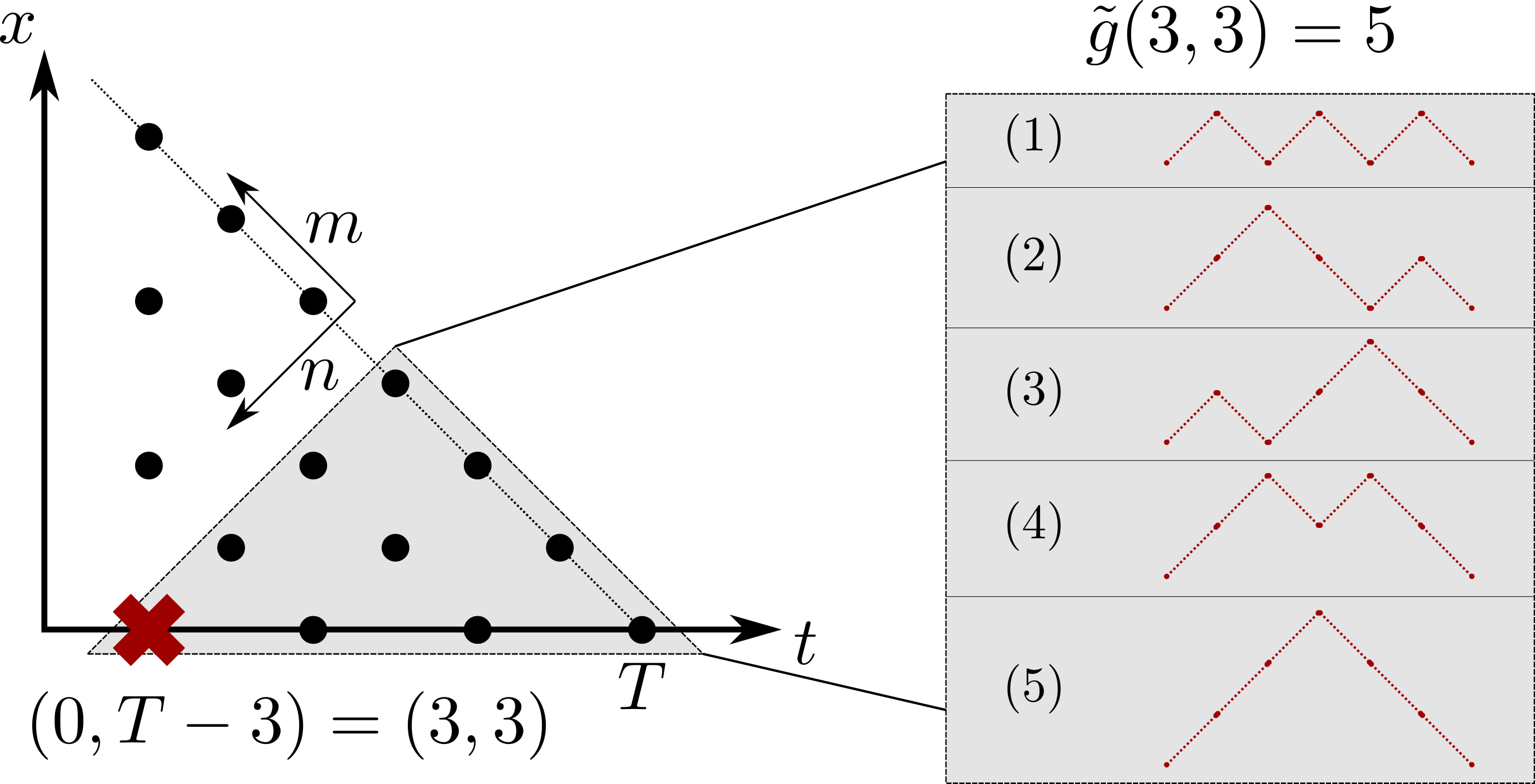}
	\end{center}
	\caption{\label{lightcone_sketch} (Left) Sketch of the backwards lightcone of points that can reach the target at $(0,T)$. The direction of the backwards lightcone coordinates are indicated by the $m$, $n$ arrows. (Right) The set of $\tilde{g}(3,3)=5$ paths leading from the point $m=3$, $n=3$ (indicated by the red cross on the left) to the target.}
\end{figure}

For the remaining components of the gauge transformation, those with $0\leq x \leq T-t$ which correspond to the probability of the remainder of the trajectory being an excursion under the original dynamics, we apply two transformations.
First, we set $g'(x,t)=2^{-t}g(x,t)$, modifying the equations to
\begin{eqnarray}
g'(x,t)=H(x+1)g'(x+1,t+1)+H(x-1)g'(x-1,t+1),
\end{eqnarray}
for $t<T-1$, with end condition $g(x,T)=1$ for all $x$ and
\begin{eqnarray}
g'(x,T-1)=\delta(x+1)+\delta(x-1).
\end{eqnarray}
Here $g'$ is interpreted as measuring the number of paths leading from the current position to the target without going below zero.
Next, we perform a coordinate transformation to backward-lightcone adapted coordinates $(m,n)$, where $m$/$n$ correspond to steps up/down going back in time (see figure \ref{lightcone_sketch}), defined by $x=m-n$ and $t=T-m-n$.
The gauge transformation in this coordinate system $\tilde{g}$ is then defined as $\tilde{g}(m,n)=g'(m-n,T-m-n)$: $\tilde{g}$ is intepreted as the number of ordered combinations of ups and downs going backwards in time for which, given any subsequence starting from the end, there are always less or equal downs than ups, i.e.\ $x\geq0$.
In these coordinates, the function $\tilde{g}$ satisfies the following set of equations
\begin{enumerate}
	\item $\tilde{g}(m,0) = 1$ for $m\geq0$,
	\item $\tilde{g}(m,1) = n$ for $m\geq1$,
	\item $\tilde{g}(m+1,n) = \tilde{g}(m+1,n-1)+\tilde{g}(m,n)$ for $1<n<m+1$,
	\item $\tilde{g}(m+1,m+1) = \tilde{g}(m+1,m)$ for $m\geq1$,
\end{enumerate}
which are precisely the equations defining Catalan's triangle, solved by 
\begin{eqnarray}
	\tilde{g}(m,n)=\frac{(m+n)!(m-n+1)}{n!(m+1)!},
\end{eqnarray}
as demonstrated in the right of figure \ref{lightcone_sketch}.
Reversing the transformations we find
\begin{eqnarray}
	g(x,t)={2^t}\tilde{g}\left(\frac{T+x-t}{2},\frac{T-x-t}{2}\right),
\end{eqnarray}
and thus
\begin{eqnarray}
	g(x,t)=\frac{1}{2^t}\frac{\left(T-t\right)!\left(x+1\right)}{\left(\frac{T-x-t}{2}\right)!\left(\frac{T+x-t+2}{2}\right)!}.
\end{eqnarray}

Finally, given this transformation, we can now calculate the transition probabilities for the optimal sampling of random walk excursions, finding
\begin{eqnarray}
	\fl P_W\left(x\pm1|x,t-1\right)
	&=\frac{1}{2}
	{2^t}\frac{\left(T-t\right)!\left(x\pm1+1\right)}{\left(\frac{T-x\mp1-t}{2}\right)!\left(\frac{T+x\pm1-t+2}{2}\right)!}
	\frac{1}{2^{t-1}}\frac{\left(\frac{T-x-t+1}{2}\right)!\left(\frac{T+x-t+3}{2}\right)!}{\left(T-t+1\right)!\left(x+1\right)}\nonumber\\
	&=\frac{1}{2}\left(1\pm\frac{1}{x+1}\right)\left(1\mp\frac{x+1\mp1}{T-t+1}\right).
\end{eqnarray}

{\renewcommand{\addtocontents}[2]{} \section{Maximum return estimation}}
\addcontentsline{toc}{section}{Appendix B~~Maximum return estimation}
\label{maximum-returns}
When training the dynamics for optimal rare trajectory sampling, the most efficient way to evaluate the current dynamics is by estimating the average return it produces.
If this average increases over time, then the model is being successfully trained.
To this end, in situations where it is available, it is useful to have an estimate for the maximum possible return over all possible transition matrices for precise evaluation of how good the model is.

This upper bound on the return can be estimates numerically by using the gauge transformations discussed in the \hyperref[exact-excursions]{Appendix A}.
First, note that since the KL divergence must be greater than 0, equation \eref{monte_carlo_dynamics_loss} immediately implies an upper bound of
\begin{eqnarray}
	\sum_{\omega_0^T}P_\theta(\omega_0^T)R(\omega_0^T)\leq\ln Z,
\end{eqnarray}
which is saturated by setting $P_\theta(x'|x,t)$ to the gauge transformed dynamics in \hyperref[exact-excursions]{Appendix A}.
We may then rewrite
\begin{eqnarray}
	Z=\sum_{\omega_0^T}W\left(\omega_{0}^T\right)P\left(\omega_{0}^T\right)=\sum_xg(x,0)p(x),
\end{eqnarray}
where $p(x)$ is the original initial state distribution.
The upper bound may then be rewritten in terms of the gauge transformation
\begin{eqnarray}
	\sum_{\omega_0^T}P_\theta(\omega_0^T)R(\omega_0^T)\leq\ln\left(\sum_xg(x,0)p(x)\right).
\end{eqnarray}

For the excursion example, this takes a particularly simple form: since the initial state distribution is $p(x)\delta_{x0}$, only a single gauge component contributes
\begin{eqnarray}
	\sum_{\omega_0^T}P_\theta(\omega_0^T)R(\omega_0^T)\leq\ln g(0,0).
\end{eqnarray}
As such, for the upper bounds in section \ref{finite-horizon} we simply need to estimate this component of the gauge transformation by numerical back-propagation of the gauge.

{\renewcommand{\addtocontents}[2]{} \section{Exact diagonalization for SCGF and optimal dynamics}}
\addcontentsline{toc}{section}{Appendix C~~Exact diagonalization for SCGF and optimal dynamics}
\label{doob-dynamics}
In order to have an accurate result for evaluation of the infinite time algorithms, we use a common technique from large deviation theory, turning the issue of finding the SCGF and optimal (Doob) sampling dynamics into one of exact diagonalization.
To this end, we first define the tilted master operator $P_s$ with components
\begin{eqnarray}
	P_s(x'|x)=P(x'|x)W_s(x,x'),
\end{eqnarray}
with the weighting parametrized by the bias $s$.
It follows simply from the definitions that the SCGF $\theta(s)$
\begin{eqnarray}
	\theta(s)
	&=\lim_{T\rightarrow\infty}\ln\left[\sum_{\omega_0^T}P(\omega_0^T)W_s(\omega_0^T)\right]\nonumber\\
	&=\lim_{T\rightarrow\infty}\ln\left[\bra{-}P_s^T\ket{P_\mathrm{ss}}\right],
\end{eqnarray}
where $\ket{P_\mathrm{ss}}$ is the steady state distribution, and thus in the infinite time limit the SCGF is simply the log of the leading eigenvalue of the matrix $P_s$.

Further to this, it is possible to calculate the optimal sampling dynamics by using this leading eigenvalue and its corresponding left eigenvector, which we label $l_s$ with components $l_s(x)$.
First, we scale the operator so that its eigenvalues are at or below zero, $P_s/e^{\theta(s)}$.
Next, we need the action of the flat state on the left of this matrix to result in zero for probability conservation: we therefore perform a basis transformation using a matrix with diagonal elements given by the components of $l_s$, finding the optimal dynamics
\begin{eqnarray}
	\tilde{P}=\frac{\mathrm{diag}(l_s)P_s\mathrm{diag}(l_s)^{-1}}{e^{\theta(s)}},
\end{eqnarray}
with the new stationary state given by component wise multiplication of the left and right eigenvectors
\begin{eqnarray}
	P^s_\mathrm{ss}(x)=l_s(x)r_s(x).
\end{eqnarray}
That this is optimal can be derived more precisely from an infinite time version of the gauge-transformation related approach of \hyperref[exact-excursions]{Appendix A} and \hyperref[maximum-returns]{Appendix B}.

\section*{References}
\bibliographystyle{unsrt}
\bibliography{dynamical_gradients}

\end{document}